\documentclass[fleqn,usenatbib]{mnras}

\usepackage[T1]{fontenc}
\usepackage{ae,aecompl}
\usepackage{soul}
\usepackage{courier}

\usepackage{graphicx}	
\usepackage{amsmath}	
\usepackage{amssymb}	

\title[MASIV IV: AGN ISS and intrinsic variability]{The MASIV Survey IV: relationship between intra-day scintillation and intrinsic variability of radio AGNs}

\author[J. Y. Koay et al.]{J. Y. Koay,$^{1,2}$\thanks{E-mail: jykoay@asiaa.sinica.edu.tw}
J.-P. Macquart,$^{3,4}$
D. L. Jauncey,$^{5,6}$
T. Pursimo,$^{7}$
M. Giroletti,$^{8}$
\newauthor H. E. Bignall,$^{9}$
J. E. J. Lovell,$^{10}$
B. J. Rickett,$^{11}$
L. Kedziora-Chudczer,$^{12,13}$
\newauthor R. Ojha,$^{14,15,16}$
C. Reynolds$^{9}$
\\
\\
$^{1}$Institute of Astronomy and Astrophysics, Academia Sinica, PO Box 23-141, Taipei 10617, Taiwan\\
$^{2}$Dark Cosmology Centre, Niels Bohr Institute, University of Copenhagen, 2100 Copenhagen \O, Denmark\\
$^{3}$ICRAR, Curtin University, Bentley, WA 6102, Australia.\\
$^{4}$ARC Centre of Excellence for All-Sky Astrophysics (CAASTRO), The University of Sydney, Sydney 2006, Australia\\
$^{5}$CSIRO Astronomy and Space Science, Epping 1710, Australia\\
$^{6}$Research School of Astronomy and Astrophysics, Australian National University, Canberra 2611, Australia\\
$^{7}$Nordic Optical Telescope, La Palma, Canary Islands 3537, Spain\\
$^{8}$INAF Istituto di Radioastronomia, via Gobetti 101, 40129 Bologna, Italy\\
$^{9}$CSIRO Astronomy and Space Science, Kensington 6151, Australia\\
$^{10}$University of Tasmania, School of Physical Sciences, Private Bag 37, Hobart 7001, Australia\\
$^{11}$University of California San Diego, La Jolla, CA 92093, USA\\
$^{12}$School of Physics, The University of New South Wales, Sydney 2052, Australia\\
$^{13}$Australian Centre for Astrobiology, The University of New South Wales, Sydney 2052, Australia\\
$^{14}$NASA Goddard Space Flight Center, 8800 Greenbelt Rd, Greenbelt, MD 20771, USA\\
$^{15}$Center for Space Science and Technology, University of Maryland, Baltimore County, 1000 Hilltop Circle, Baltimore, MD 21250, USA\\
$^{16}$Department of Physics, The Catholic University of America, 620 Michigan Ave NE, Washington, DC 20064, USA\\
}

\date{Accepted XXX. Received YYY; in original form ZZZ}

\pubyear{2017}

\begin{document}
\label{firstpage}
\pagerange{\pageref{firstpage}--\pageref{lastpage}}
\maketitle

\begin{abstract}
We investigate the relationship between 5\,GHz interstellar scintillation (ISS) and 15\,GHz intrinsic variability of compact, radio-selected AGNs drawn from the Microarcsecond Scintillation-Induced Variability (MASIV) Survey and the Owens Valley Radio Observatory (OVRO) blazar monitoring program. We discover that the strongest scintillators at 5\,GHz (modulation index, $m_5 \geq 0.02$) all exhibit strong 15\,GHz intrinsic variability ($m_{15} \geq 0.1$). This relationship can be attributed mainly to the mutual dependence of intrinsic variability and ISS amplitudes on radio core compactness at $\sim 100\, \mu$as scales, and to a lesser extent, on their mutual dependences on source flux density, arcsec-scale core dominance and redshift. However, not all sources displaying strong intrinsic variations show high amplitude scintillation, since ISS is also strongly dependent on Galactic line-of-sight scattering properties. This observed relationship between intrinsic variability and ISS highlights the importance of optimizing the observing frequency, cadence, timespan and sky coverage of future radio variability surveys, such that these two effects can be better distinguished to study the underlying physics. For the full MASIV sample, we find that \textit{Fermi}-detected gamma-ray loud sources exhibit significantly higher 5\,GHz ISS amplitudes than gamma-ray quiet sources. This relationship is weaker than the known correlation between gamma-ray loudness and the 15\,GHz variability amplitudes, most likely due to jet opacity effects.  
\end{abstract}




\section{Introduction}\label{intro}

The variability of compact Active Galactic Nuclei (AGNs) at radio wavelengths can originate from intrinsic phenomena such as the propagation of shocks along jets \citep{hughesetal85, hovattaetal08}, jet precession \citep[e.g.,][]{kudryavtsevaetal2011}, variable accretion rates due to disk instabilities  \citep[e.g.,][]{linshields86} or tidal disruption events \citep[][and references therein]{donnarummaetal15}. Studying such intrinsic radio variations of AGNs, in relation to the variability observed at optical to gamma-ray frequencies, potentially provides important clues on the physical structure of AGNs on scales often unresolved by existing telescopes, plus the physics of supermassive black hole accretion and jets. 

At $\sim 5$ to 8\,GHz, AGN variability on timescales of hours to days is dominated by scintillation due to scattering in the ionized interstellar medium (ISM) of our Galaxy \citep{heeschenrickett87, rickett90}. The Micro-Arcsecond Scintillation Induced Variability (MASIV) Survey \citep{lovelletal03} of $\sim 500$ compact, flat-spectrum AGNs at 5\,GHz enabled the first large statistical study of AGN intra and inter-day variability. 58\% of the MASIV Survey sources were found to be variable in at least one of the four epochs of observations \citep{lovelletal08}; the strong Galactic dependence of the variability amplitudes comfirms that the flux variations are predominantly due to interstellar scintillation (ISS). Although ISS introduces unwanted noise into studies of intrinsic AGN radio variability, its sensitivity to source sizes at microarcsecond ($\mu$as) scales makes it an excellent probe of compact structures in radio AGNs \citep[e.g.,][]{macquartjauncey02,macquartetal13,rickettetal02}, as well as the turbulent properties of the ISM \citep{armstrongetal95}.

The recent revival of interest in time-domain astronomy across all wavelengths has generated a number of variability studies using new and archival data on existing radio telescopes such as the Karl G. Jansky Very Large Array \citep[VLA,][]{ofekfrail11, hodgeetal13}, the Allen Telescope Array \citep[ATA,][]{boweretal11}, the Australia Telescope Compact Array \citep[ATCA,][]{belletal15,hancocketal16} and the Murchison Widefield Array \citep[MWA,][]{belletal14}. While single-field studies are improving our understanding of the sky density of variable sources and its dependence on the flux-limits of these surveys \citep[e.g.,][]{mooleyetal16}, there is only a cursory understanding of the origin of the observed AGN variability in these studies. In the meantime, new surveys have been planned for upcoming radio telescopes, including the Variable and Slow Transients (VAST) Survey \citep{murphyetal13} on the Australia Square Kilometre Array Pathfinder (ASKAP), and eventually the Square Kilometre Array \citep{bignalletal15}.

To maximize the science output from these ongoing and upcoming radio variability surveys, there is a need to better understand the origin of and physical processes responsible for the observed flux variations. One key aspect is to differentiate between intrinsic and extrinsic (ISS-induced) variability of radio AGNs \citep{bignalletal15}. At what timescales and at which frequencies does one effect begin to dominate over the other? Do either one of these two different modes of variability exhibit an exclusive dependence on another property (e.g., Galactic latitude, source size, source flux density), such that the dominant mode of variability can be distinguished from the other in a large sample of variable sources? What is the relationship between ISS and intrinsic variability, i.e., are the most intrinsically variable sources also expected to be the strongest scintillators? There are no existing studies aimed at understanding the relationship between intrinsic and extrinsic radio variability of AGNs for a large sample of sources. Yet, such studies are crucial for informing the design of future variability surveys, such as in optimizing the observing frequency and cadence, depending on whether the aim is to study ISS or intrinsic radio variations. If the goal is to study both phenomena, then future surveys will need to be designed such that these two effects can be clearly distinguished in the data.

In this paper, we present a pilot study examining the relationship between inter-day ISS and longer timescale intrinsic variability. Since our ISS data are drawn from the MASIV Survey, we briefly summarize the observations, characterization of the variability amplitudes, and main results of the survey in Section~\ref{MASIV}. In Section~\ref{intrinsicvar}, we discuss the relationship between MASIV ISS amplitudes and the strengths of the predominantly intrinsic 15\,GHz flux variations measured over 4 years by the Owens Valley Radio Observatory (OVRO) blazar monitoring program \citep{richardsetal11,richardsetal14}, for a subset of sources observed in both programs. Using published VLBI data \citep{pushkarevkovalev15}, we argue that the statistically significant correlation between the observed 5\,GHz and 15\,GHz variability amplitudes can be attributed to their mutual dependence on the angular sizes of the sub-milliarcsecond radio core components (Section~\ref{massize}). In Section~\ref{mutualothers}, we also investigate the mutual dependence of the 5\,GHz and 15\,GHz variability amplitudes on source spectral indices, flux densities, redshifts, and gamma-ray loudness, and whether they may contribute to the observed correlation between ISS and intrinsic variability amplitudes. Finally in Section~\ref{summary}, we summarise our main results and discuss their implications for future variability surveys. We adopt the following cosmology: $\Omega_m = 0.3$, $\Omega_\Lambda = 0.70$, and $H_0 = 70\,{\rm km\,s^{-1}\,Mpc^{-1}}$. 

\section{The MASIV Survey}\label{MASIV}

In the MASIV Survey \citep{lovelletal03,lovelletal08}, the 5\,GHz flux densities of 475 sources were monitored with the VLA over 4 epochs, each of 3 or 4 day duration, spaced $\sim$4 months apart. These observations were conducted in the years 2002 and 2003. Each source was observed for 1 minute every 2 hours while above an elevation of $15^{\circ}$.

The MASIV survey sources were selected to be compact and to have flat spectral indices ($\alpha \geq -0.3$, for $S \propto \nu^{\alpha}$), based on non-coeval flux measurements from the 1.4\,GHz NRAO VLA Sky Survey \citep{condonetal98}, plus the Jodrell Bank VLA Astrometric Survey \citep{patnaiketal92,browneetal98,wilkinsonetal98} and Cosmic Lens All-Sky Survey \citep{myersetal95} at 8.4\,GHz. The MASIV source sample is flux-limited; about half the sources have flux densities of typically $\sim 100$\,mJy (classified as `radio-weak'), while the other half have flux densities of $\sim 1$\,Jy (classified as `radio-strong').

Variability amplitudes at timescales of $\tau = 2$\,hrs to 2 days (in 2\,hr increments) were characterized using the flux normalized structure function, $D(\tau)$, derived from the observed lightcurves. The following simple model \citep{lovelletal08},
\begin{equation}\label{Dmodel}
D(\tau) = 2m^2\dfrac{\tau}{\tau + \tau_{\rm char}}, 
\end{equation}
was fit to the observed $D(\tau)$ values of each source. This model assumes that the variations are a stationary stochastic process, such that $D(\tau)$ slowly rises with increasing values of $\tau$ and saturates at an amplitude of $2m^2$, which is related to the modulation index, $m$. $\tau_{\rm char}$ is the characteristic variability timescale where  $D(\tau)$ reaches half its saturation amplitude. The model values of $D(\tau = \rm 2d)$ were selected as the representative values of the intra/inter-day variability amplitudes of the MASIV Survey sources.

58\% of the MASIV sources were found to exhibit 2 to 10\% rms variations in at least one of the 4 epochs. $D(\rm 2d)$ was found to have a strong dependence on Galactic latitudes, confirming that it is dominated by ISS. ISS amplitudes are also strongly dependent on the intrinsic properties of the AGNs, due to their high sensitivity to source compactness. These include source flux density \citep{lovelletal08}, redshift \citep{lovelletal08,koayetal12,pursimoetal13}, milliarcsecond (mas) scale radio core dominance \citep{ojhaetal04}, radio spectral index \citep{koayetal11}, and optical spectroscopic classification of AGN type \citep{pursimoetal13}. Some of these results are further elucidated below as they become relevant to our discussion. Otherwise, we refer the reader to the above papers for more details.

\subsection{This Study}\label{masivthisstudy}

For our present study, we remove 21 MASIV sources that have either negative values of $D(\rm 2d)$ or that do not have well-determined $D(\rm 2d)$ due to non-convergent model fits. We also remove another 30 sources for which the errors in $D(\rm 2d)$ are two orders of magnitude larger than $D(\rm 2d)$. This leaves us with 424 sources.

For the purposes of this study, we characterize the amplitude of ISS at 5\,GHz using the modulation index, $m_{5}$, derived from $D(\rm 2d)$ as follows \citep{lovelletal08}: 
\begin{equation}\label{mod} 
m = \sqrt{D(\rm 2d)/2} \,\,.
\end{equation}
This assumes that at 2-day timescales, $D(\tau)$ has already saturated. Converting to the modulation indices allows for like-for-like comparisons with the published modulation indices used in the characterization of the variability amplitudes of sources in the OVRO monitoring program, which we describe next.

\section{AGN variability at 5\,GHz and 15\,GHz}\label{intrinsicvar}

\subsection{The 15\,GHz OVRO blazar monitoring program}\label{ovro}

\citet{richardsetal14} in their paper present the 15\,GHz variability amplitudes of $\sim$1500 blazars based on 4 years of monitoring by the OVRO 40\,m telescope. Each source was observed at a cadence of two flux measurements a week between 2008 January to 2011 December. The original sample of 1158 sources monitored by OVRO \citep{richardsetal11} were selected from the Candidate Gamma-Ray Blazar Survey \citep[CGRaBS,][]{healeyetal08}; these CGRaBS sources were selected to have spectral indices, radio flux densities and X-ray flux densities similar to sources with EGRET gamma-ray detections, such that they would have a high chance of being detected in gamma-rays by \textit{Fermi}. After the release of the \textit{Fermi}-LAT catalogue, an additional $\sim 400$ gamma-ray detected sources from the First LAT AGN Catalogue \citep[1LAC,][]{abdoetal10} were added to the OVRO monitoring sample. For this present paper, we focus only on the radio-selected CGRaBS sources from the \citet{richardsetal11} sample, since we are interested in understanding the variability characteristics of radio-selected AGNs. We ignore the additional gamma-ray selected 1LAC sources in their sample.

To characterize the variability amplitudes of the OVRO lightcurves, the `intrinsic modulation index', described in \citet{richardsetal11}, is used. This estimate uses the maximum likelihood method to determine the standard deviation of the flux densities of each source lightcurve, normalized by the mean flux density. 

A total of 178 of the remaining 424 sources in our MASIV sample are also in the original radio-selected OVRO 15\,GHz monitoring sample. We refer to this sample of 178 sources, the intersection between the MASIV and \citet{richardsetal14} source samples, as the M $\cap$ R sample; we use this sample for our analyses in this section. The list of 178 sources is presented in Table~\ref{datatable}, together with the variability amplitudes of the sources and other source properties. Of this M $\cap$ R sample, 173 sources are radio-strong ($S_5 \geq 0.3$\,Jy), while only 5 sources are radio-weak with $S_5 < 0.3$\,Jy. Our results for this paper therefore pertain mainly to bright, high flux density radio AGNs. 

\begin{table*}
	\centering
	\caption{List of M $\cap$ R sample of sources and their variability characteristics, flux densities, redshifts,  line-of-sight H$\alpha$ intensities, and VLBI core sizes. The full version of this table is available online in electronic form.
		\label{datatable}}
	\begin{tabular}{l c c c c c c c c c c c c c}
		\hline
		\hline
    	Name & $m_5$ & Ep  & $S_5$ & $m_{15}$ & $S_{15}$ & $I_{\alpha}$ & $z$ & $\theta_{2}$ & $\theta_{5}$ & $\theta_{8}$ & $\theta_{15}$ & $\theta_{24}$ & $\theta_{43}$ \\
        (J2000) &  & & (Jy) &  & (Jy) & R &  & (mas) & (mas) & (mas) & (mas) & (mas) & (mas) \\
		\hline
J0005+3820 & 0.015$\pm$0.003 & 1 & 0.52 & 0.164$^{+0.008}_{-0.007}$ & 0.53 &  2.9 & 0.23 & 0.56 & -& 0.27 & 0.13 & -& -\\ 
J0010+1724 & 0.007$\pm$0.003 & 0 & 0.82 & 0.062$^{+0.003}_{-0.003}$ & 0.49 &  0.8 & 1.60 & 0.88 & -& 0.52 & -& -& -\\ 
J0017+8135 & 0.010$\pm$0.003 & 0 & 0.92 & 0.043$^{+0.003}_{-0.003}$ & 0.92 &  2.2 & 3.39 & 0.81 & 0.63 & 0.27 & 0.31 & 0.26 & -\\ 
J0019+2021 & 0.016$\pm$0.004 & 1 & 1.11 & 0.112$^{+0.005}_{-0.005}$ & 0.60 &  0.7 & - & - & 1.16 & 0.37 & - & 0.10 & 0.08 \\ 
J0038+4137 & 0.017$\pm$0.004 & 2 & 0.53 & 0.047$^{+0.002}_{-0.002}$ & 0.53 &  2.6 & 1.35 & 1.14 & 1.63 & 0.26 & 0.23 & - & - \\ 
J0056+1625 & 0.026$\pm$0.007 & 3 & 0.24 & 0.321$^{+0.016}_{-0.014}$ & 0.33 &  0.8 & 0.21 & 1.27 & - & 0.10 & - & - & - \\ 
J0057+3021 & 0.011$\pm$0.003 & 0 & 0.68 & 0.054$^{+0.003}_{-0.003}$ & 0.86 &  1.2 & 0.02 & 1.61 & 1.02 & 0.59 & 0.30 & - & -\\ 
		\hline
	\end{tabular}
	\begin{flushleft}
	\textbf{Column notes:} (2) 5\,GHz modulation index \citep{lovelletal08}; (3) Number of epochs variable in the MASIV survey \citep{lovelletal08}; (4) 5\,GHz flux density \citep{lovelletal08}; (5) 15\,GHz modulation index \citep{richardsetal14}; (6) 15\,GHz flux density \citep{richardsetal14}; (7) line-of-sight H$\alpha$ intensity \citep{haffneretal03}; (8) source redshift \citep[][Pursimo et al. in prep]{pursimoetal13}; (9-14) VLBI core sizes at 2, 5, 8, 15, 24, 43\,GHz \citep{pushkarevkovalev15}.  \\
	\end{flushleft}
\end{table*}

\subsection{ISS or intrinsic variability?}\label{intextvar}

We first examine if the 15\,GHz and 5\,GHz modulation indices exhibit significant dependences on line-of-sight Galactic H$\alpha$ intensities, to determine if they are dominated by ISS or intrinsic variability. The H$\alpha$ intensity, obtained from the Wisconsin H-Alpha Mapper (WHAM) survey \citep{haffneretal03}, is a measure of the integral of the squared electron density along the line of sight; it therefore functions as a proxy for the Galactic line-of-sight scattering strength towards each source. The intra and inter-day variability amplitudes of compact AGNs from the MASIV survey are known to depend significantly on their line-of-sight H$\alpha$ intensities, both at 5\,GHz \citep{lovelletal08} and 8\,GHz \citep{koayetal11}. This demonstrates that the flux variations at these frequencies on timescales of a few days are dominated by ISS. At 15\,GHz, the variability on timescales of months to years is expected to be dominated by intrinsic variations, and was assumed to be so by \citet{richardsetal14}. We note however, that ISS has been observed at 15\,GHz for some sources, such as QSO 1156+295 \citep{savolainenkovalev08}. 

We confirm that the significant dependence of the 5\,GHz modulation indices, $m_5$, on the line-of-sight H$\alpha$ intensities, $I_{\alpha}$, holds true for our M $\cap$ R sample of 178 sources; the two-tailed, non-parametric Spearman correlation test gives a $p$-value of $2.28 \times 10^{-3}$ (where the significance level is selected to be $\alpha = 0.05$). The $p$-value is the probability of observing such a correlation (or stronger) in our data purely by chance, assuming the null hypothesis that there is no correlation is true. When performing the Spearman partial correlation test on a smaller sample of 157 sources with known redshift, we still find a significant correlation ($p = 5.15 \times 10^{-4}$) between $m_5$ and $I_{\alpha}$ after controlling for flux density, spectral index, redshift, and the 15\,GHz modulation index, $m_{15}$. A summary of the results of all our Spearman rank correlation tests between different variables and for different source subsamples are shown in Table~\ref{spearmantable}. The results of our Spearman partial correlation tests are summarized in Table~\ref{partialcorrtable}. A brief description of these tests are provided in Appendix~\ref{corrstat}. We note that we also performed Kendall-Tau correlation tests alongside the Spearman correlation tests. Since the results of both tests are fully consistent, we do not present the Kendall-Tau test results. 

\begin{table*}
	\centering
	\caption{Spearman rank correlation coeffficients, $r_s$, and corresponding $p$-values between pairs of parameters and for various source samples.
		\label{spearmantable}}
	\begin{tabular}{l c c c c c c}
		\hline
		\hline
		Parameter 1 & Parameter 2 & Source sample & No. of sources & $r_s$ & $p$-value & Significant?\\
		\hline
		$m_5$   &	$I_{\alpha}$	& 	M $\cap$ R  & 178 & 	0.227		& $2.28 \times 10^{-3}$	& Y\\
		$m_{15}$	 &  $I_{\alpha}$	&  M $\cap$ R  & 178 &	$-$0.018	& 	$8.07 \times 10^{-1}$	& N\\
		\hline
		$m_5$	&  $m_{15}$	& M $\cap$ R  & 178	& 0.275 & 	$2.08 \times 10^{-4}$ & Y\\
		&  	& 	M $\cap$ R (gamma-ray loud)	& 75 & 	0.241	& $3.71 \times 10^{-2}$	& Y	\\
		&  	& 	M $\cap$ R (gamma-ray quiet)	&  103 &	0.285	& 	$3.54 \times 10^{-3}$	& Y		 \\
		&  	&  M $\cap$ R	($I_{\alpha} \geq 0.6$ R)	& 147	& 0.315	& 	$1.03 \times 10^{-4}$	& Y	 \\
		&  	& 	M $\cap$ R ($I_{\alpha} < 0.6$ R)	&  31 & 	0.384	& 	$3.30 \times 10^{-2}$	& Y	 \\
		&  	& 	M $\cap$ R (FSRQ)	&  124 & 	0.200	& 	$2.59\times 10^{-2}$  & Y\\
		&  	& 	M $\cap$ R (BL Lac)	&  36 & 	0.237	& 	$1.65\times 10^{-1}$	& N	 \\
		&  	& 	M $\cap$ R ($S_{5} \geq 0.8$  Jy)	& 102	& 0.293	& 	$2.79 \times 10^{-3}$  & Y\\
		&  	& 		M $\cap$ R ($S_{5} < 0.8$ Jy)	& 76	& 0.264	& $2.11 \times 10^{-2}$	& Y \\
		&  	& 	M $\cap$ R ($z \geq 2$)	& 25	& 0.343	& $9.31 \times 10^{-2}$ & N  \\
		&  	& 	M $\cap$ R ($z < 2$)	& 132	& 0.200	& 	$2.17 \times 10^{-2}$ & Y	 \\
		\hline
		$m_5$	&  $\theta_{\rm 2}$	& 	M $\cap$ PK2	& 246	& $-$0.227	& 	$3.17 \times 10^{-4}$	& Y \\
		$m_5$	&  $\theta_{\rm 5}$	& 	M $\cap$ PK5	& 210	& $-$0.264	& 	$1.07 \times 10^{-4}$	 & Y\\
		$m_5$	&  $\theta_{\rm 8}$	& 	M $\cap$ PK8	& 260	& $-$0.385	& 	$1.37 \times 10^{-10}$	& Y \\
		$m_5$	&  $\theta_{\rm 15}$	& 	M $\cap$ PK15	& 129	& $-$0.353	& 	$4.02 \times 10^{-5}$	& Y \\
		$m_5$	&  $\theta_{\rm 24}$	& 	M $\cap$ PK24	& 96	& $-$0.272	& 	$7.31 \times 10^{-3}$	& Y \\
		$m_5$	&  $\theta_{\rm 43}$	& 	M $\cap$ PK43	& 50	& $-$0.284	& 	$4.56 \times 10^{-2}$	& Y \\
		\hline
		$m_{15}$	&  $\theta_{\rm 2}$	& 	M $\cap$ R $\cap$ PK2	& 164	& $-$0.234	& 	$2.59\times 10^{-3}$	& Y \\
		$m_{15}$	&  $\theta_{\rm 5}$	& 	M $\cap$ R $\cap$ PK5	& 113	& $-$0.201	& 	$3.27 \times 10^{-2}$	& Y \\
		$m_{15}$	&  $\theta_{\rm 8}$	& 	M $\cap$ R $\cap$ PK8	& 171	& $-$0.434	& 	$3.07\times 10^{-9}$	& Y \\
		$m_{15}$	&  $\theta_{\rm 15}$	& 	M $\cap$ R $\cap$ PK15	& 117	& $-$0.468	& 	$1.00  \times 10^{-7}$	& Y \\
		$m_{15}$	&  $\theta_{\rm 24}$	& 	M $\cap$ R $\cap$ PK24	& 83	& $-$0.158	& 	$1.53 \times 10^{-1}$	& N \\
		$m_{15}$	&  $\theta_{\rm 43}$	& 	M $\cap$ R $\cap$ PK43	& 46	& $-$0.258	& 	$8.34 \times 10^{-2}$ & N	 \\
		\hline
		$m_{5}$	&  $S_{5}$	& 	M $\cap$ R & 178	& $-$0.315	& 	$1.84\times 10^{-5}$ & Y \\
		$m_{15}$	&  $S_{5}$	& 	M $\cap$ R 	& 178	& $-$0.021	& $7.83 \times 10^{-1}$ & N \\
		$m_{15}$	&  $S_{15}$	& 	M $\cap$ R 	& 178	& 0.171	& $2.26\times 10^{-2}$ & Y \\
		\hline
		$m_{5}$	&  $\alpha_{5}^{15}$	& 	M $\cap$ R 	& 178 & 0.197 & 	$8.41\times 10^{-3}$ & Y	 \\
		$m_{15}$	&  $\alpha_{5}^{15}$	& 	M $\cap$ R 	& 178 & 0.330 & 	$7.06\times 10^{-6}$ & Y	 \\
		\hline
		$m_{5}$	&  $z$	& 	M $\cap$ R (known $z$) 	& 157	& $-$0.164	& 	$3.96\times 10^{-2}$	& Y \\
		$m_{15}$	&  $z$	& 	M $\cap$ R (known $z$) & 157	& $-$0.354& 	$5.34\times 10^{-6}$ 	& Y \\
		\hline
	\end{tabular}
\end{table*}

\begin{table*}
	\centering
	\caption{Spearman rank partial correlation coeffficients, $r_s$, and corresponding $p$-values between pairs of variables while controlling for other variables. Strikethrough text in the third column indicates variables not used as control variables (see text in Section~\ref{mutualflux}).
		\label{partialcorrtable}}
	\begin{tabular}{l c c c c c c c}
		\hline
		\hline
		Variable 1 & Variable 2 & Control Variable(s) & Source sample & No. of sources & $r_s$ & $p$-value & Significant?\\
		\hline
		$m_5$   &	$I_{\alpha}$	& - & M $\cap$ R (known $z$) & 157 & 	0.275	& $5.03 \times 10^{-4}$	& Y\\
		$m_5$   &	$I_{\alpha}$	& $m_{15}$, $S_5$, $\alpha_{5}^{15}$, $z$ & M $\cap$ R (known $z$) & 157 & 	0.278	& $5.15 \times 10^{-4}$	& Y\\
		$m_{15}$	 &  $I_{\alpha}$	& - & M $\cap$ R (known $z$)  & 157 &	$-$0.049	&  $5.40 \times 10^{-1}$	& N \\
		$m_{15}$	 &  $I_{\alpha}$	& $m_5$, $S_{15}$, $\alpha_{5}^{15}$, $z$ & M $\cap$ R (known $z$)  & 157 &	$-$0.141	& 	$8.31 \times 10^{-2}$ & N \\
		\hline
		$m_5$	&  $m_{15}$	&  - & M $\cap$ R $\cap$ PK8 & 171	& 0.276 & 	$2.62 \times 10^{-4}$ & Y \\
		$m_5$	&  $m_{15}$	&  $\theta_{8}$ & M $\cap$ R $\cap$ PK8 & 171	& 0.137 & 	$7.55  \times 10^{-2}$ & N \\
		$m_5$	&  $m_{15}$	&  - & M $\cap$ R $\cap$ PK15 & 117	& 0.276 & 	$2.60 \times 10^{-3}$ & Y\\
		$m_5$	&  $m_{15}$	&  $\theta_{15}$ & M $\cap$ R $\cap$ PK15 & 117	& 0.152 & 	$1.03 \times 10^{-1}$ & N\\
		\hline
		$m_5$	&  $m_{15}$	&  - & M $\cap$ R $\cap$ PK8 (known $z$) & 151	& 0.236 & 	$3.50 \times 10^{-3}$ & Y \\
		$m_5$	&  $m_{15}$	&  $\theta_{8}$, $S_5$, $S_{15}$, $\alpha_{5}^{15}$, $z$ & M $\cap$ R $\cap$ PK8 (known $z$) & 151	& 0.105 & 	$2.08 \times 10^{-1}$ & N\\
		$m_5$	&  $m_{15}$	&  \textbf{\st{$\theta_{8}$}}, $S_5$, $S_{15}$, $\alpha_{5}^{15}$, $z$ & M $\cap$ R $\cap$ PK8 (known $z$) & 151	& 0.192 & 	$1.99 \times 10^{-2}$ & Y\\
		$m_5$	&  $m_{15}$	&  $\theta_{8}$, \st{$S_5$}, $S_{15}$, $\alpha_{5}^{15}$, $z$ & M $\cap$ R $\cap$ PK8 (known $z$) & 151	& 0.092 & 	$2.70 \times 10^{-1}$ & N\\
		$m_5$	&  $m_{15}$	&  $\theta_{8}$, $S_5$, \st{$S_{15}$}, $\alpha_{5}^{15}$, $z$ & M $\cap$ R $\cap$ PK8 (known $z$) & 151	& 0.102 & 	$2.21 \times 10^{-1}$ & N\\
		$m_5$	&  $m_{15}$	&  $\theta_{8}$, $S_5$, $S_{15}$, \st{$\alpha_{5}^{15}$}, $z$ & M $\cap$ R $\cap$ PK8 (known $z$) & 151	& 0.094 & 	$2.58 \times 10^{-1}$ & N\\
		$m_5$	&  $m_{15}$	&  $\theta_{8}$, $S_5$, $S_{15}$, $\alpha_{5}^{15}$, \st{$z$} & M $\cap$ R $\cap$ PK8 (known $z$) & 151	& 0.149 & 	$7.11 \times 10^{-2}$ & N\\
		\hline
	\end{tabular}
\end{table*}

On the other hand, we find no significant correlation ($p$-value of 0.807) between the 15\,GHz modulation indices, $m_{15}$, and the line-of-sight H$\alpha$ intensities of these 178 sources. This holds true for the smaller sample of 157 sources with known $z$, after controlling for flux density, spectral index, redshift, and $m_5$ (Table~\ref{partialcorrtable}). This demonstrates that while the 5\,GHz variability of our sample is dominated by ISS, $m_{15}$ is most likely dominated by intrinsic variations. 

The fact that the 15\,GHz variability amplitudes are characterized using the modulation indices, which are dominated by the largest flux excursions, and are derived based on the 4-year lightcurves (unlike $m_5$ that is derived from the structure function at 2-day timescales), also increases the likelihood that $m_{15}$ is dominated by the larger amplitude and longer timescale (months and years) intrinsic variations as opposed to ISS. 

\subsection{Relationship between 5\,GHz and 15\,GHz variability amplitudes}\label{intextrel}

\begin{figure*}
\begin{center}
\includegraphics[width=\textwidth]{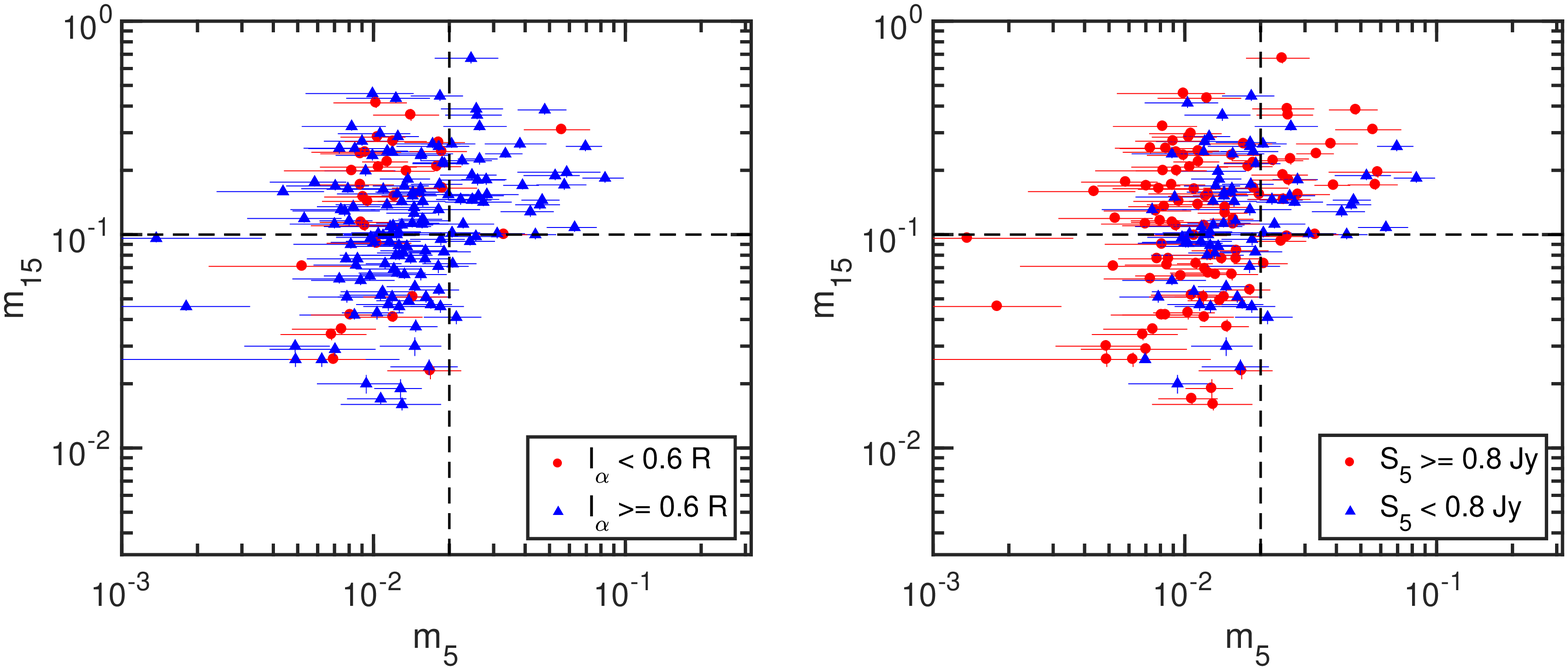}
\end{center}
\caption{{15\,GHz modulation index from the OVRO monitoring program vs the 5\,GHz modulation index obtained from the MASIV survey for the M $\cap$ R sample of 178 sources, with sources separated based on their line of sight H$\alpha$ intensities (left) and flux densities (right). The vertical and horizontal lines denote $m_{5} = 0.02$ and $m_{15} = 0.1$, respectively. There is a statistically significant correlation between the strength of the predominantly intrinsic variations at 15\,GHz and the amplitude of ISS at 5\,GHz. \label{varcomp}}}
\end{figure*}

We find a statistically significant correlation between $m_{15}$ and $m_{5}$ for our sample of sources (Figure~\ref{varcomp}). The Spearman test gives a correlation coefficient of $r_s = 0.275$ and a $p$-value of $2.08 \times 10^{-4}$. This is in spite of the fact that the MASIV observations and the first year of the OVRO blazar monitoring campaign are spaced $\sim$6 years apart. 

With the exception of the BL Lac objects and sources at $z \geq 2$ (possibly due to their small sample sizes), this correlation between intrinsic variability and ISS amplitudes remains significant for all the other sub-categories of sources, grouped based on their gamma-ray loudness, line-of-sight H$\alpha$ intensity, flux density and redshift (see Table~\ref{spearmantable}). 

While the $p$-values show that the correlations are statistically significant, the low correlation coefficients indicate that these correlations are weak. This is not surprising since the AGN variability amplitudes (regardless if they are intrinsic or due to ISS) are dependent on many other factors (discussed further in Sections~\ref{massize} and \ref{mutualothers}).

\begin{figure}
	\begin{center}
		\includegraphics[width=\columnwidth]{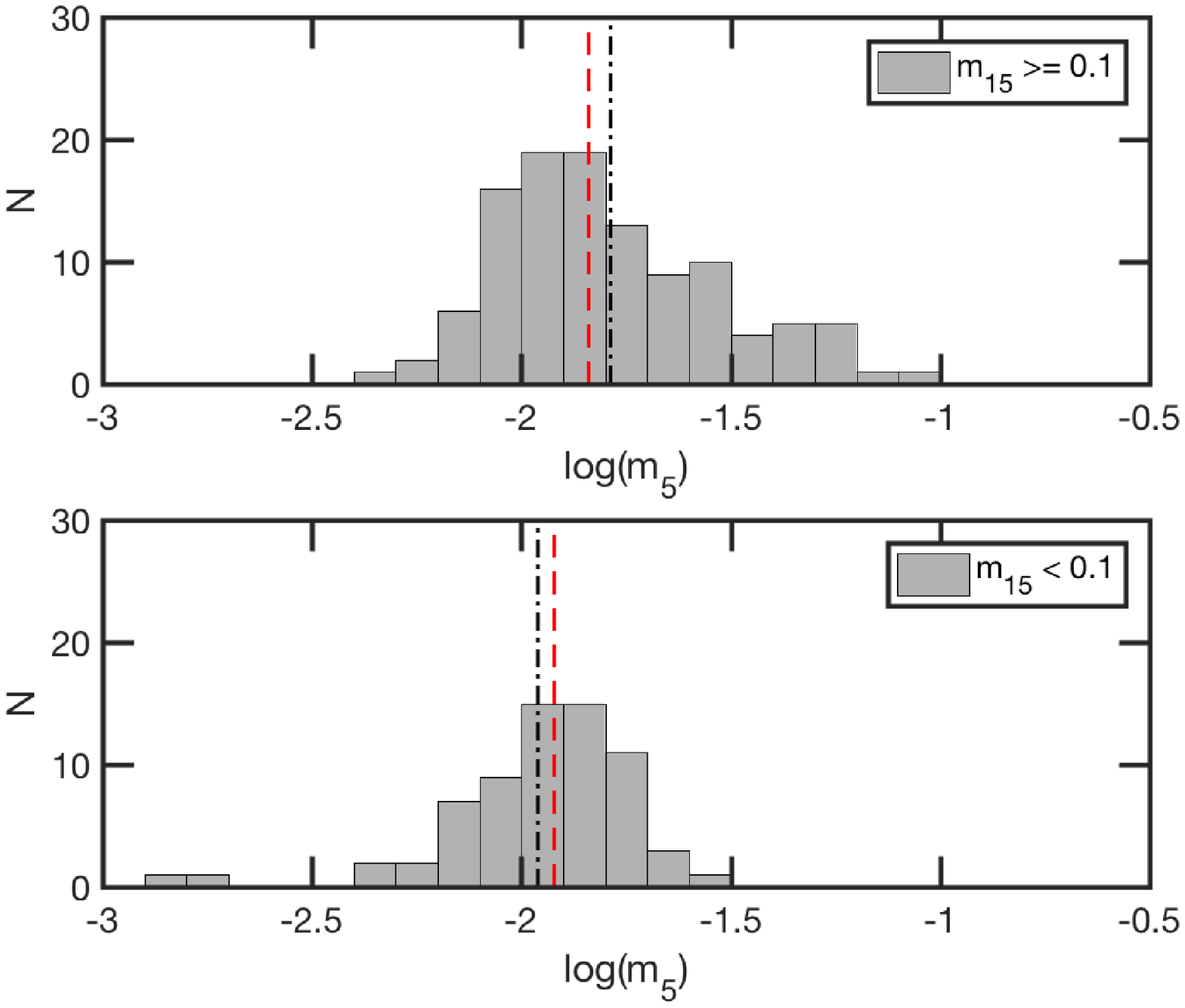}
	\end{center}
	\caption{{Histograms showing the distributions of $m_{5}$ for sources with strong ($m_{15} \geq 0.1$, top) and weak ($m_{15} < 0.1$, bottom) 15\,GHz modulation indices. The dashed (red) and dash-dotted (black) vertical lines show the median and mean values of $m_{5}$, respectively. \label{m5dist}}}
\end{figure}

\begin{figure}
	\begin{center}
		\includegraphics[width=\columnwidth]{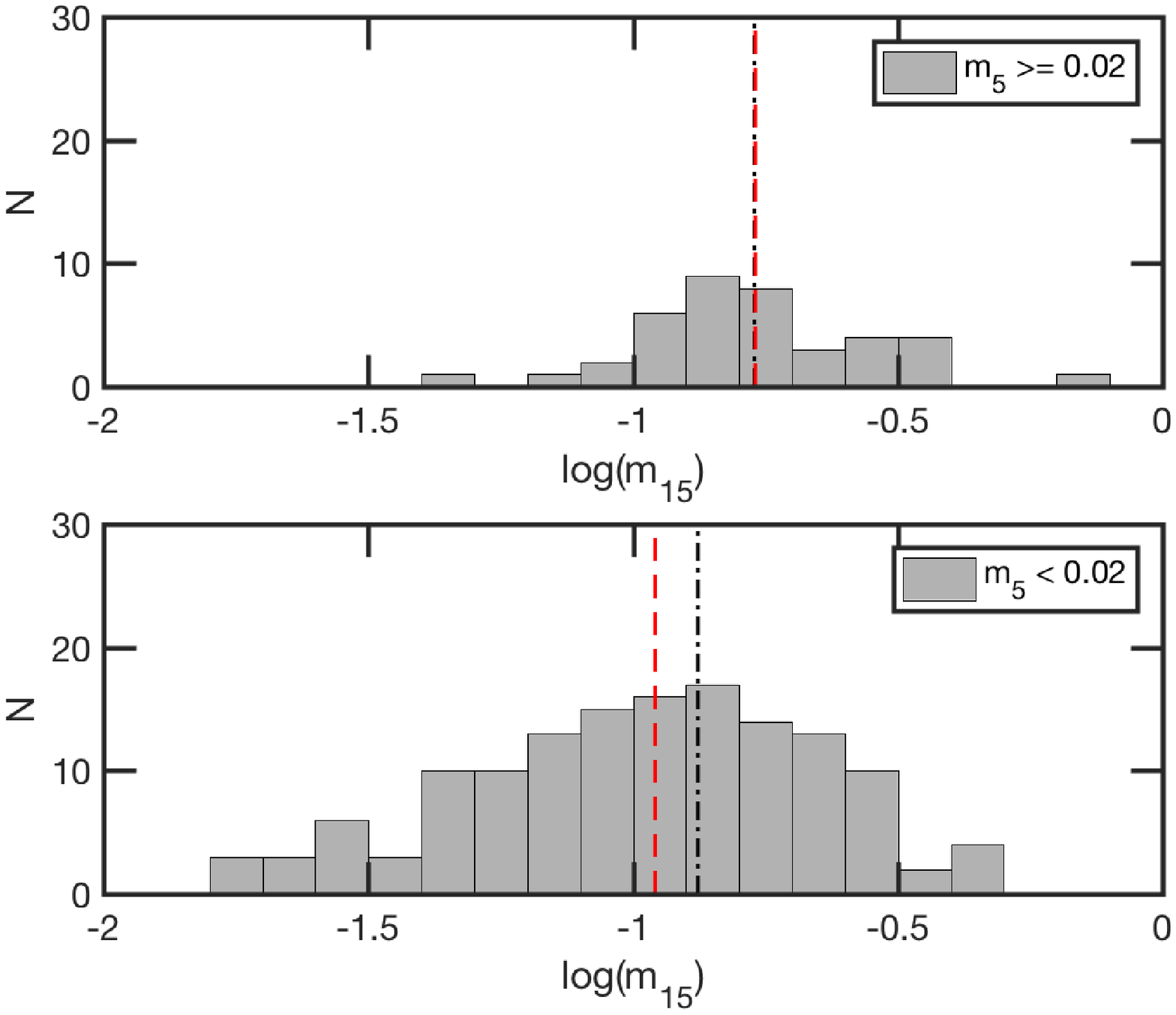}
	\end{center}
	\caption{{Histograms showing the distributions of $m_{15}$ for sources with strong ($m_{5} \geq 0.02$, top) and weak ($m_{5} < 0.02$, bottom) 5\,GHz modulation indices. The dashed (red) and dash-dotted (black) vertical lines show the median and mean values of $m_{15}$, respectively. \label{m15dist}}}
\end{figure}

As an alternative, we compare the distributions of $m_5$ for sources showing large 15\,GHz variability ($m_{15} \geq 0.1$) with that of sources showing weak 15\,GHz variability ($m_{15} < 0.1$) in Figure~\ref{m5dist}. The Kolmogorov-Smirnov (K-S) test rejects the null hypothesis that the distributions of $m_5 (m_{15}  \geq 0.1)$ and $m_5 (m_{15}  < 0.1)$ are drawn from the same parent population (with a $p$-value of $1.37 \times 10^{-3}$). We also show the distributions of $m_{15}$ for sources exhibiting large ($m_{5} \geq 0.02$) and small ($m_{5} < 0.02$) amplitude variations at 5\,GHz Figure~\ref{m15dist}. The K-S test also finds, at a significant level, that the distributions of $m_{15}(m_{5} \geq 0.02)$ and $m_{15}(m_{5} < 0.02)$ are not drawn from the same parent population ($p = 3.25 \times 10^{-4}$).

\subsection{Interpretation as interdependence between ISS and intrinsic variability}\label{intextcorr}

We argue that the observed inter-relationship between $m_{5}$ and $m_{15}$ as described above is best interpreted in the framework where $m_{5}$ is dominated by ISS while $m_{15}$ is characterized predominantly by intrinsic AGN variability. We thus adopt this interpretation here and for the rest of the paper. We consider (and argue against) alternative interpretations in Appendix~\ref{altexp}, using ISS models and Monte-Carlo simulations. 

It can be seen from Figures~\ref{varcomp} and \ref{m5dist} that sources with lower intrinsic variability ($m_{15} < 0.1$) tend to have lower levels of ISS ($m_{5} \lesssim 0.02$). Sources with ISS amplitudes of $m_{5} \gtrsim 0.02$ also typically show strong intrinsic variations of $m_{15} \geq 0.1$. However, there is a population of sources that exhibit strong intrinsic variability ($m_{15} \geq 0.1$), but exhibit low ISS amplitudes ($m_{5} < 0.02$). In other words: (i) the highest-amplitude scintillators all show strong intrinsic variability (as shown in Figure~\ref{m15dist}), but (ii) not all sources that show strong intrinsic variations are large-amplitude scintillators. We discuss these statements separately here:

\begin{enumerate}
	\item \ul{the highest amplitude scintillators all show strong intrinsic variability}. At 5\,GHz, the angular size of a source must be $\lesssim 135 \, \mu$as for it to scintillate at amplitudes of $m_{5} \geq 0.02$ on a timescale of 2 days. We derived this estimate based on the \citet{goodmannarayan06} fitting formula for ISS, assuming fiducial values of the scattering screen distance (500\,pc), scattering screen velocity (30\,km\,s$^{-1}$), and transition frequency between weak and strong scattering (5\,GHz). We also assume that the power spectrum of the electron density fluctuations in the ISM follow a power law consistent with Kolmogorov turbulence; this is observationally motivated \citep{armstrongetal95}. At the mean redshift of $z \sim 1$ for our M $\cap$ R sample, sources exhibiting ISS at amplitudes of $m_{5} \geq 0.02$ are thus expected to have components with linear sizes of $\lesssim 1$\,pc. As a zeroth order estimate, such a component is sufficiently compact to exhibit significant intrinsic variability on a timescale of $\lesssim 4$ years, based on light travel time arguments. Sources with lower-level ISS amplitudes will (on average) have angular sizes $\gtrsim 135 \, \mu$as, corresponding to linear sizes of $\gtrsim 1$\,pc, such that any significant intrinsic variations will be observed on much longer timescales and thus will be missed in the first 4 years of the OVRO monitoring program. For sources at $z \gtrsim 1$, the apparent intrinsic variability timescales of a 1\,pc diameter component may exceed 4 years due to time dilation effects. However, this effect is typically offset by the time compression arising from the Doppler-boosted, relativistically beamed emission in these sources. 
	
	\item \ul{Not all sources that show strong intrinsic variations are large-amplitude scintillators.} This can be explained by the fact that ISS amplitudes are dependent not only on the intrinsic source properties (i.e. source compactness), but also on the properties of the scattering material in the line-of-sight. Selecting only sources with large intrinsic variability, $m_{15} \geq 0.1$, we present the line-of-sight H$\alpha$ intensity distributions separately for sources with low ISS amplitudes  ($m_5 < 0.02$) and high ISS amplitudes ($m_5 \geq 0.02$) in Figure~\ref{varialphahist}. We find that sources that show lower level ISS tend to have lower line-of-sight H$\alpha$ intensities compared to sources that show higher amplitude ISS, within this sample of sources with strong intrinsic variations (see also Figure~\ref{varcomp}, left panel). The K-S test rejects the null hypothesis that the H$\alpha$ intensities of these two sub-samples are drawn from the same parent population, with a $p$-value of $0.029$. Some of these sources with high levels of intrinsic variability are thus likely scintillating only very weakly due to relatively lower levels of scattering in the line of sight. 
	
\end{enumerate}

\begin{figure}
	\begin{center}
		\includegraphics[width=\columnwidth]{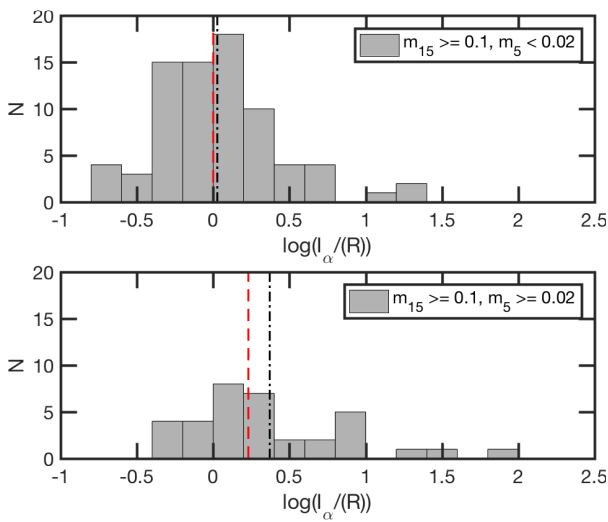}
	\end{center}
	\caption{{Histograms showing the distributions of the line-of-sight H-$\alpha$ intensities ($I_{\alpha}$) for sources with strong intrinsic variations ($m_{15} \geq 0.1)$), separated into those with weak ISS ($m_5 < 0.02$, top) and strong ISS ($m_5 \geq 0.02$, bottom). The dashed (red) and dash-dotted (black) vertical lines show the median and mean values of $I_{\alpha}$, respectively. Due to the logarithmic binning, one of the sources in the top panel with a negative $I_{\alpha}$ has been place in the lowest $I_{\alpha}$ bin. \label{varIalphahist}}}
\end{figure}

\subsection{ISS intermittency and intrinsic variability}\label{intextepoch}

\begin{figure}
	\begin{center}
		\includegraphics[width=\columnwidth]{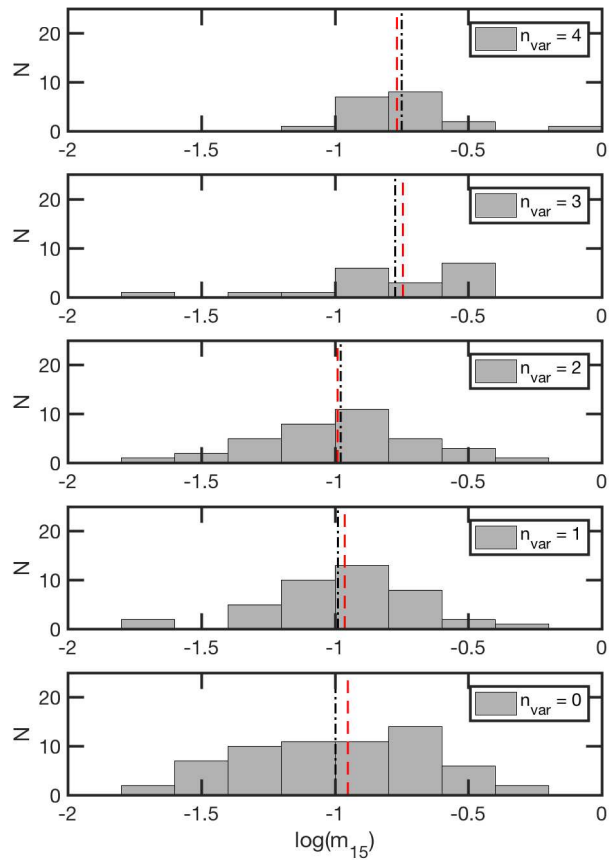}
	\end{center}
	\caption{{Histograms showing the distributions of the 15\,GHz modulation indices, with sources categorized based on the number of epochs, $n_{\rm var}$, in the MASIV survey in which they exhibited ISS. The dashed (red) and dash-dotted (black) vertical lines show the median and mean values of $m_{15}$, respectively. \label{varepochs}}}
\end{figure}

We also find that sources that are variable in 3 or all 4 of the MASIV epochs at 5\,GHz are more likely to have stronger variability amplitudes at 15\,GHz. The distributions of $m_{15}$ for source sub-samples categorized based on the number of epochs in which they were found to scintillate during the MASIV survey, are shown in Figure~\ref{varepochs}. The K-S test finds that the distribution of $m_{15}$ for sources that scintillate in 3 and 4 epochs differs significantly ($p$-value of $5.42 \times 10^{-6}$) from that of sources that scintillate in 2 or less epochs. 

The intermittency of AGN ISS is thought to arise due to either (i) factors intrinsic to the AGN itself, i.e., the transient nature of compact scintillating components, or (ii) extrinsic factors, i.e., the fluctuating turbulence and structure of the ISM. The observed intermittency of ISS in archetypal scintillating AGNs such as PKS\,0405$-$385 \citep{kedziora-chudczer06} and J1819+3845 \citep{koayetal11b,debruynmacquart15} appears to be consistent with line-of-sight structural variations of the local ISM. However, our result demonstrating the relationship between intrinsic variability amplitudes and the persistence of ISS suggests that intrinsic factors are also important.

\begin{figure}
	\begin{center}
		\includegraphics[width=0.85\columnwidth]{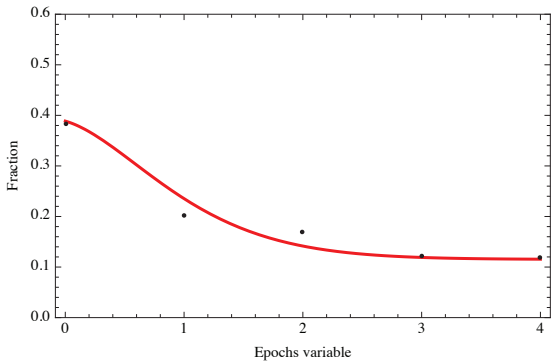}
	\end{center}
	\caption{{The fraction of sources variable in 0 to 4 epochs, inclusive of the perennial sources. The red curve shows a fit to the Poisson distribution. \label{burst}}}
\end{figure}

We examine a simple, bursting model in which the incidence of variability (i.e., the presence of a compact source component due, e.g., to a flare) is treated as a Poisson process. We use the incidence of intra-day variability in the four MASIV epochs \citep{lovelletal08}, shown in Table~\ref{varincidence}, to determine the rate associated with this process, making the simple assumption that each of the four epochs are sampled independently from the distribution (i.e., that the fading time between bursts is less than the inter-epoch separation). However, we account for this assumption post hoc by incorporating into the model a fraction of sources that are classed as `perennially' compact, which we estimate as the fraction of sources that were observed to be variable in all four epochs of the MASIV survey. We estimate the fraction of sources variable in 0 to 3 epochs, after removing the fraction of perennial variables, and list them in the third column of Table~\ref{varincidence}. We fit to these fractions the following equation:
\begin{equation}
f(n) = \frac{e^{-\mu} \mu^n}{n!},
\end{equation}
from which we determine the rate parameter of the Poisson distribution as $\mu=0.39$. $n$ is the number of epochs in which a burst occurs. On the basis of this fit, the total fraction of sources variable across $n$ well-separated epochs is:
\begin{equation}
f = \frac{51}{424} + \left( 1- 5 \times \frac{51}{424} \right ) \frac{e^{-\mu} \mu^n}{n!},
\end{equation}
shown as the red curve in Figure~\ref{burst}, with the perennials included. 

\begin{table}
	\centering
	\caption{Fraction of sources observed to be variable in $n = 0, 1, 2 ...$ MASIV epochs (with perennials removed) and the probability ($f(n \geq 1)$) that these sources will be observed to be variable at some future interval, for rate parameters of $\mu = 0, 1, 2 ...$
		\label{varincidence}}
	\begin{tabular}{lcc} 
		\hline 
		\hline
		Epochs & no. of &  fraction with \\ 
		variable & sources & perennials removed \\
		\hline
		0	& 163  & 0.66 \\ 
		1	& 86 & 0.21 \\ 
		2	& 72& 0.12 \\ 
		3	& 52 & 0.01 \\ 
		4	& 51 & 0.00 \\ 
		\hline
		total & 424 & 1.000  \\
		\hline 
	\end{tabular} 
\end{table}

The fraction of sources that will exhibit variability at some future interval (which we loosely define as undergoing at least one burst per year, $n \geq$ 1) is then given by:
\begin{align}\label{burstfutureeq}
f(n \geq 1) &= 1 - f(n = 0)  \,\,.
\end{align}
where we obtain $f(n \geq 1) = 0.61$. This is regardless of the number of epochs in which the source was found to be variable in the MASIV survey, since the two intervals are independent. This simple model predicts that $m_{15}$ should not exhibit any dependence on the persistence of ISS in the MASIV survey. This discrepancy between model and observational data reveals that (i) the flux variations at two separate epochs (within the MASIV Survey and between the MASIV observations and the OVRO program) are NOT independent of each other, e.g., some bursts extend over more than one epoch; and/or (ii) the burst rates of different sources are not equal, i.e., some sources are more inclined than others to exhibit bursts.

One possible explanation for this observed relationship between $m_{15}$ and the persistency of ISS in the MASIV Survey, is that the sources found to be variable over 3 to 4 MASIV epochs contain long-lived compact components that continually scintillate, making these objects more likely to still be scintillating and intrinsically variable 6 to 10 years later during the OVRO monitoring program. For a 100\,$\mu$as source ($\sim 1$\,pc at a typical redshift of 1), the light-crossing time is $\sim 3$ years. This explanation is thus valid for sources with low Lorentz factors where the intrinsic source structures are not expected to vary in size on timescales less than a few years.

However, compact blazars such as those in our sample, often exhibiting brightness temperatures exceeding the $5 \times 10^{10}\,$K equipartition limit \citep{readhead94, lahteenmakietal99, lahteenmakivaltaoja99}, are known from VLBI observations to have Doppler boosting factors ranging from $\delta \sim 5$ to 30 \citep[e.g.,][]{lahteenmakivaltaoja99,hovattaetal09}, with median values of $\delta \sim$15 and 6 for FSRQs and BL Lac objects respectively. These sources are thus capable of exhibiting significant structural variations on timescales of a few months. For such highly relativistic sources, the alternative explanation is that those exhibiting strong intrinsic variations during the OVRO monitoring program are likely to undergo more frequent flaring episodes in the observer's frame (possibly due to time compression), and are thus more likely to be ejecting new compact scintillating components on a more frequent basis. This increases the likelihood of detecting 5\,GHz ISS in these sources during the MASIV survey campaign. Such sources will need to eject a new compact (scintillating and intrinsically variable) component on timescales much shorter than (i) the 4\,month interval between each MASIV epoch, and/or (ii) the time it takes for these components to expand significantly and dissipate. For sources that scintillate in 2 or fewer MASIV epochs, the compact scintillating components may expand and dissipate on timescales less than the typical time interval between successive outbursts/flares responsible for launching new compact components. 

\section{Dependence of ISS and intrinsic variability on VLBI core sizes}\label{massize}

We propose that the inter-dependence between the 5\,GHz MASIV ISS amplitudes and the OVRO 15\,GHz intrinsic variations stem mainly from the dependence of both on the intrinsic sizes of the AGN radio cores. We now examine the relationships between the variability amplitudes at both frequencies and the source sizes at mas and sub-mas scales, as derived from VLBI observations. 

\subsection{Published VLBI core sizes}\label{VLBIdata}

\citet{pushkarevkovalev15} compiled a sample of more than 3000 sources that have VLBI observations over multiple epochs and multiple frequencies. These sources are mainly drawn from the Radio Fundamental Catalogue (RFC) comprising all sources observed with VLBI for astrometry and geodesy programmes. They also include data from major VLBI surveys such as the VLBA Calibrator Survey \cite[e.g.,][]{beasleyetal02,petrovetal05,kovalevetal07}, the VLBI Imaging and Polarimetry Survey \citep{helmboldtetal07,petrovtaylor11}, and the VLBA 2-cm Survey \cite{kellermannetal98,kellermannetal04}, among others. For their sample of sources, they fit two circular Gaussian components to the self-calibrated visibilities for each source, then select the most compact component as the core component for which they determine the FWHM angular sizes. For sources in which multi-epoch data are available, they present the median value of the angular size. The uncertainties in the core size estimates were not presented. We refer readers to the original paper by \citet{pushkarevkovalev15} for more details on the source samples and methodology. 

\subsection{Dependence of ISS on source size}\label{ISStheta}

Of the 424 MASIV sources, 246, 210, 260, 129, 96, and 50 sources have estimates of the 2\,GHz, 5\,GHz, 8\,GHz, 15\,GHz, 24\,GHz and 43\,GHz FWHM source angular sizes repectively. We refer to these overlapping samples at each frequency as M $\cap$ PK2, M $\cap$ PK5, M $\cap$ PK8, M $\cap$ PK15, M $\cap$ PK24 and M $\cap$ PK43 respectively.

\begin{figure*}
	\begin{center}
		\includegraphics[width=\textwidth]{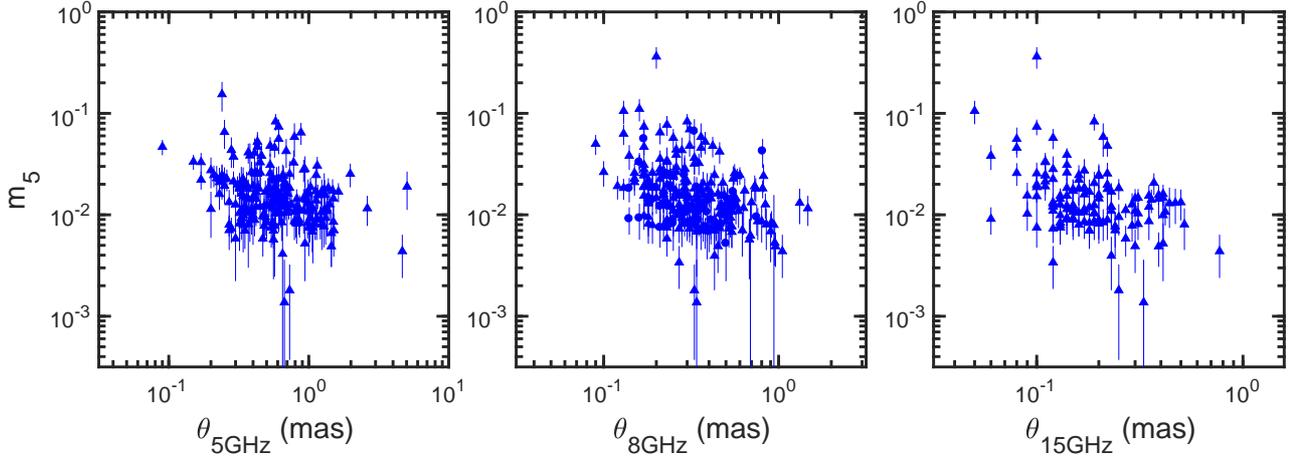}
	\end{center}
	\caption{{Relationship between the 5\,GHz modulation indices, $m_{5}$, and the FWHM angular sizes of the core component at 5\,GHz, 8\,GHz, and 15\,GHz, derived by \citet{pushkarevkovalev15} using VLBA data.  \label{varcorrsize_m5}}}
\end{figure*}

We find that the ISS amplitudes, $m_5$, show a significant dependence on the FWHM source sizes measured at 2 to 43\,GHz, based on the Spearman correlation test. The $r_s$ correlation coefficients and $p$-values are presented in Table~\ref{spearmantable}, while plots of $m_5$ against core sizes at 5, 8, and 15\,GHz are shown in Figure~\ref{varcorrsize_m5}. 

We note that the correlations are stronger, with higher $r_s$, when comparing $m_5$ with source angular sizes at 8\,GHz to 43\,GHz instead of at 5\,GHz, despite the ISS observations being conducted at 5\,GHz. This is likely to be due to the higher angular resolutions at higher frequencies, such that the scales being probed are approaching the expected angular scales ($\lesssim 150\,\mu$as) of the scintillating components. Even though different regions along the jet are being probed at different frequencies due to opacity effects, \citet{pushkarevkovalev15} demonstrate that there is a core size-frequency dependence of $\theta \propto \nu^{-k}$ for their sample of sources; the distribution of $k$ peaks at a value of 1 for sources at latitudes $|b| > 10^{\circ}$ (where ISM scatter broadening effects do not dominate), consistent with a conical jet model \citep{blandfordkonigl79}. 

\citet{ojhaetal04} demonstrated that scintillating sources are more core-dominated than non-scintillating sources at mas scales for a sub-sample of 237 MASIV Survey sources. Our results expand upon this work and show that ISS amplitudes are also dependent on the FWHM core sizes derived from VLBI measurements at (sub-)mas resolution. We predict that ISS amplitudes will show an even stronger dependence on the radio core sizes and core dominances at the tens of $\mu$as scales probed by \textit{RadioAstron} on Earth-space baselines.

\subsection{Dependence of intrinsic variability on source size}\label{intvartheta}

\begin{figure*}
	\begin{center}
		\includegraphics[width=\textwidth]{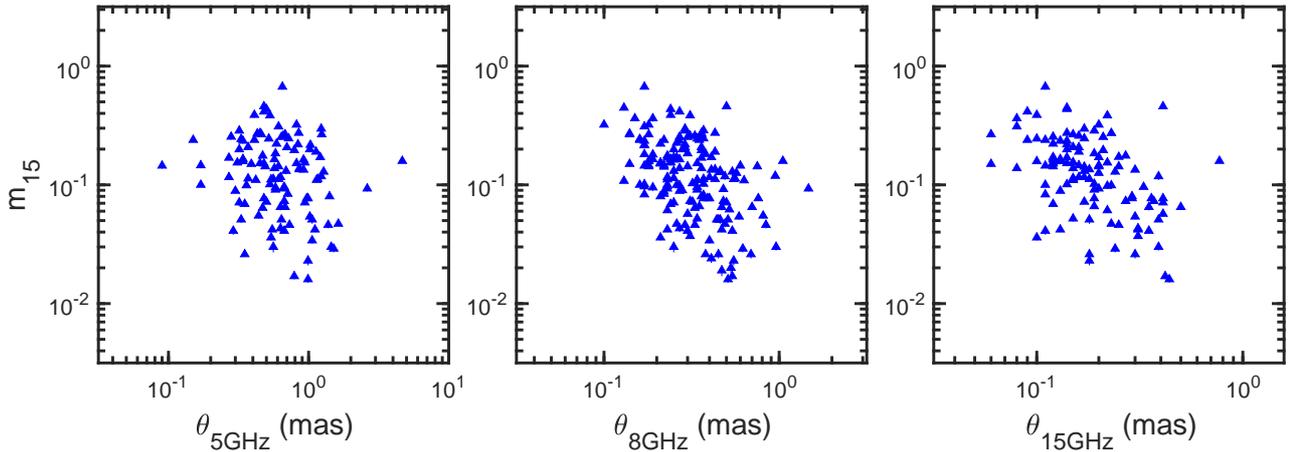}
	\end{center}
	\caption{{Relationship between the 15\,GHz modulation indices, $m_{15}$, and the FWHM angular sizes of the core component at 5\,GHz, 8\,GHz, and 15\,GHz, derived by \citet{pushkarevkovalev15} using VLBA data. \label{varcorrsize_m15}}}
\end{figure*}

Selecting only sources that are found in both the MASIV and \citet{richardsetal14} sample (M $\cap$ R), we find that there are VLBI source size measurements for 164, 113, 171, 117, 83, and 46 sources at 2\,GHz, 5\,GHz, 8\,GHz, 15\,GHz, 24\,GHz, and 43\,GHz respectively. We refer to these samples as M $\cap$ R $\cap$ PK2, M $\cap$ R $\cap$ PK5, M $\cap$ R $\cap$ PK8, M $\cap$ R $\cap$ PK15, M $\cap$ R $\cap$ PK24, and M $\cap$ R $\cap$ PK43 respectively. Plots of $m_{15}$ vs FWHM core sizes measured at 5, 8, and 15\,GHz are shown in Figure~\ref{varcorrsize_m15}. The Spearman test reveals that $m_{15}$ is significantly dependent on the core sizes measured at 2\,GHz to 15\,GHz (Table~\ref{spearmantable}). Not surprisingly, the correlation coefficient is highest for $\theta_{\rm 15}$, since the intrinsic variability amplitude is measured at 15\,GHz, where the VLBI observations are probing the 100\,$\mu$as ($\sim 0.8$ parsec at $z \sim 1$) scale structures of the core, very likely the components exhibiting the $\lesssim$ 4-year timescale intrinsic variability in the source. 

\subsection{Effect of sub-mas core sizes on the relationship between $m_5$ and $m_{15}$}\label{mutualtheta}

We argue that the observed significant relationship between ISS and intrinsic variability amplitudes arises mainly due to the dependence of both on the intrinsic sizes of the AGN cores. The size of the emitting region places a lower limit on the observed timescale of intrinsic variations, which in turn affects the amplitude of variations that can be observed within a finite timespan of a monitoring program. On the other hand, ISS is also highly sensitive to source angular sizes. The scintillation patterns produced by the scattering of waves originating from different regions of an extended source are smeared out, decreasing the ISS amplitudes (analogous to the suppression of atmospheric scintillation of planets relative to that of stars).

\begin{figure*}
	\begin{center}
		\includegraphics[width=\textwidth]{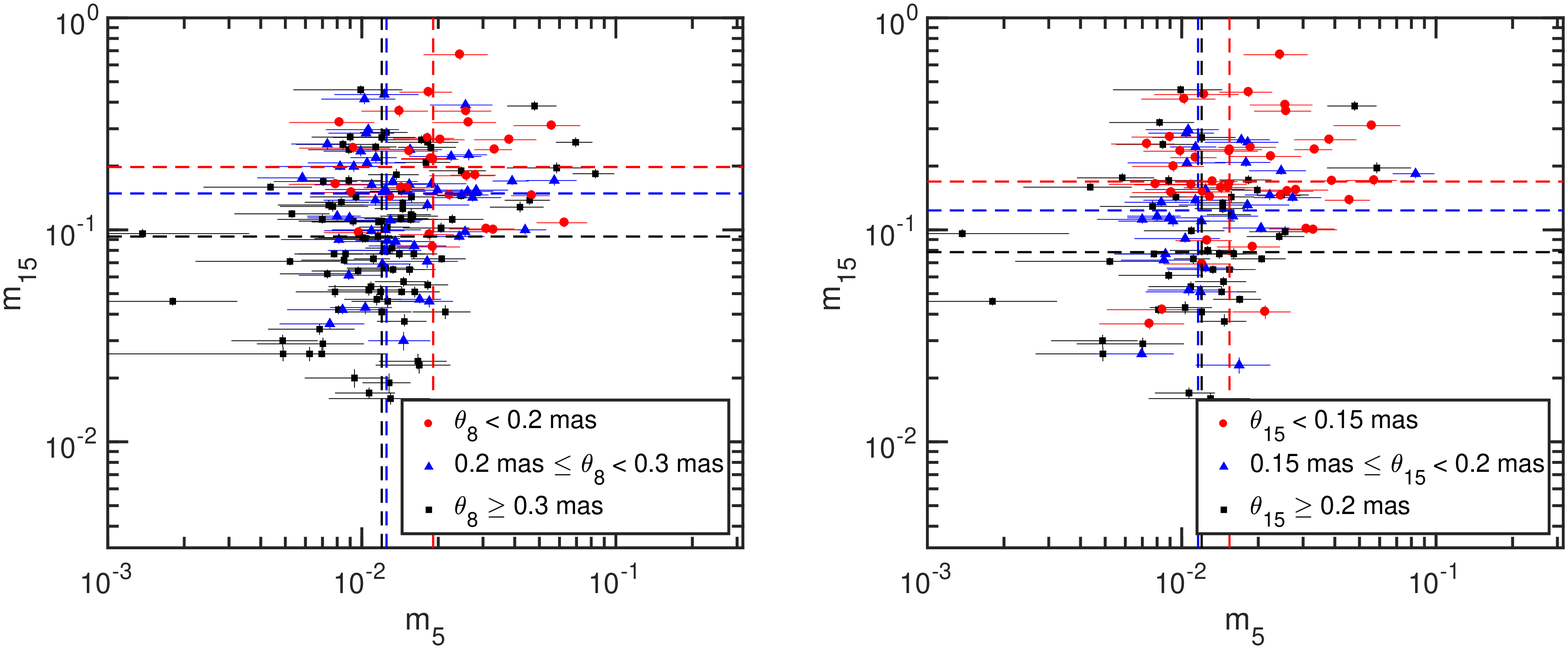}
	\end{center}
	\caption{{15\,GHz modulation index from the OVRO monitoring program vs the 5\,GHz modulation index obtained from the MASIV survey for the M $\cap$ R $\cap$ PK8 sample (left), and M $\cap$ R $\cap$ PK15 sample (right), with sources binned based on their mas core sizes measured at 8\,GHz (left) and 15\,GHz (right). The vertical and horizontal lines denote median values of $m_{5}$ and $m_{15}$ for each of the core size bins.  \label{varcompsizebin}}}
\end{figure*}

This mutual dependence is demonstrated in Figure~\ref{varcompsizebin}. Sources with the most compact sub-mas radio cores ($\theta_{8} < 0.2$\,mas and $\theta_{15} < 0.15$\,mas, red circles) have higher median values of $m_{15}$ and $m_{5}$ compared to sources with more extended cores. We selected $\theta_{8}$ and  $\theta_{15}$  as representative characterizations of the core sizes since they have the highest correlation coefficients with respect to both $m_5$ and $m_{15}$. 

When controlling for $\theta_8$ and $\theta_{15}$ separately using the Spearman partial correlation test, we find that the correlation between $m_5$ and $m_{15}$ becomes insignificant (with $p$-values of $7.55 \times 10^{-2}$ and $1.03 \times 10^{-1}$ respectively). The $p$-values are now at least 2 orders of magnitude lower when compared to that of our Spearman tests in Table~\ref{spearmantable}. This reduction in the $p$-values cannot be attributed to the reduction in the sample size alone, when selecting only sources with VLBI measurements of $\theta_5$ and $\theta_8$. In Table~\ref{partialcorrtable}, we show also the results of the Spearman correlation test between $m_5$ and $m_{15}$ with no controlled variables, but with the same sample size used for the partial correlation test (as the control sample), for comparisons. For comparable samples, the lower $p$-values and correlation coefficients when controlling for $\theta$ confirms the influence of source size in the $m_5$ and $m_{15}$ correlation.

\section{Dependence of ISS and intrinsic variability on other intrinsic properties}\label{mutualothers}

We now examine if the relationship between $m_{5}$ and $m_{15}$ also arises due to their mutual dependences on other intrinsic source properties, i.e., spectral indices, flux densities, redshift, and gamma-ray loudness. 

\subsection{Source flux density, spectral index and redshift}\label{mutualflux}

As mentioned in Section~\ref{intro}, ISS amplitudes show a strong dependence on source flux density \citep{lovelletal08,pursimoetal13}. The angular size of a source can be modeled as a function of its flux density, $S$, and the intrinsic brightness temperature, $T_{\rm b}$, as follows:
\begin{equation}\label{thetaTb} 
\theta = \sqrt{\dfrac{\lambda^{2}(1+z)S}{2\pi k \delta T_{\rm b}}} 
\end{equation}
where $\lambda$ is the observing wavelength, $z$ is the source redshift, $k$ is the Boltzmann constant, and $\delta$ is the Doppler boosting factor. Therefore, if these compact extragalactic sources are limited in brightness temperature, either due to the inverse-Compton catastrophe \citep{kellermannpauliny-toth69} or energy equipartition between the magnetic fields and the electrons \citep{readhead94, lahteenmakietal99}, the source angular sizes are expected to scale as $\theta \propto \sqrt{S}$. This in turn leads to weaker levels of ISS in higher flux density sources. This flux density dependence of ISS remains significant in our M $\cap$ R source sample; the Spearman correlation test between $m_5$ and the 5\,GHz flux density ($S_5$) gives a $p$-value of $1.84\times 10^{-5}$. This can also be seen in the right panel of Figure~\ref{varcomp}, where the weaker $S_{5} < 0.8$\,Jy sources have modulation indices that are typically higher than that of the $S_{5} \gtrsim 0.8$\,Jy sources. We observe a similar dependence of $m_{15}$ on $S_{15}$ ($p$-value of $2.26 \times 10^{-2}$), as expected for soures with smaller angular size being more intrinsically variable.  

It is also known that both ISS and intrinsic variability amplitudes exhibit a significant redshift dependence \citep{lovelletal08,richardsetal11}. For ISS, this redshift dependence is consistent with the cosmological $\theta \propto \sqrt{1+z}$ effect expected for a flux limited and brightness temperature-limited sample of sources \citep{koayetal12,pursimoetal13}. On the other hand, the redshift dependence of the 15\,GHz intrinsic variations has been attributed to time-dilation of the variability timescales of high redshift sources \citep{richardsetal11}. For our M $\cap$ R source sample, this redshift dependence is still significant for both $m_5$ and $m_{15}$ (Table~\ref{spearmantable}). For this paper, we have adopted the source redshifts from \citet{pursimoetal13}, as well as new, spectroscopic redshifts that have yet to be published (Pursimo et al., in prep). 

Despite the non-coeval MASIV and OVRO flux measurements, $m_5$ and $m_{15}$ still show significant correlations with the spectral index derived from the mean flux densities at 5\,GHz and 15\,GHz (Table~\ref{spearmantable}). Since the modulation indices at both frequencies are normalized by the total measured flux densities of the source, sources whose emission are dominated by that of the variable component will have larger modulation indices. Sources with inverted and flat spectra (and are thus more core-dominated at arcsecond scales) exhibit larger amplitude ISS and intrinsic variations. 

Although these flux, redshift and spectral index dependences likely also contribute to the observed correlation between $m_5$ and $m_{15}$, the $\sim 100 \, \mu$as core size is the dominant factor in determining the variability amplitudes at both 5 and 15\,GHz. While $m_5$ and $m_{15}$ exhibit mutual dependences on the source flux densities, spectral indices and redshift, their correlation coefficients are slightly lower when compared to the correlation coefficients between $m_5$ (as well as $m_{15}$) and the core sizes, particularly at 8\,GHz and 15\,GHz. To examine this further, we perform the Spearman partial correlation test between $m_5$ and $m_{15}$, controlling for all 5 parameters $\theta_8$, $S_5$, $S_{15}$, $\alpha_{5}^{15}$, and $z$, for the M $\cap$ R $\cap$ PK8 sample of sources with known redshifts (with a total of 151 sources). We obtain a correlation coefficient of $0.105$ and $p$-value of $2.08 \times 10^{-1}$, which indicates that the correlation between $m_5$ and $m_{15}$ is no longer significant. This is in contrast to when no other variables are controlled ($r_s = 0.236$,  $p = 3.50 \times 10^{-3}$) for the same sample of 151 sources. We then repeat the test for each instance in which one of the variables are removed as a controlled variable. The correlation coefficients and $p$-values are shown in Table~\ref{partialcorrtable}, where in the third column of the table, each variable struck off with a horizontal line indicates that they were not controlled for in the test. We find that the correlation between $m_5$ and $m_{15}$ remains insignificant ($p > 0.05$) when $S_5$, $S_{15}$, $\alpha_{5}^{15}$, or $z$ are removed as controlled variables. However, the correlation between $m_5$ and $m_{15}$ becomes significant when only $\theta_8$ is not controlled for. 

\subsection{Gamma-ray loudness}\label{mutualgammaray}

We now discuss the dependence of both modes of radio variability on the gamma-ray loudness of blazars. \citet{richardsetal11} found, for their sample of 1158 blazars monitored by the OVRO telescope over 2 years, that sources with a \textit{Fermi}-detected counterpart in the 1LAC catalogue have significantly higher intrinsic variability amplitudes at 15\,GHz than that of gamma-ray quiet sources. This result is further confirmed in their updated work with 4 years of monitoring data; \citet{richardsetal14} report that for their radio-selected sample of CGRaBS AGNs, gamma-ray loud sources that have a clean association in the Second LAT AGN Catalog (2LAC) have significantly higher $m_{15}$ than gamma-ray quiet AGNs. 

\begin{figure*}
	\begin{center}
		\includegraphics[width=\textwidth]{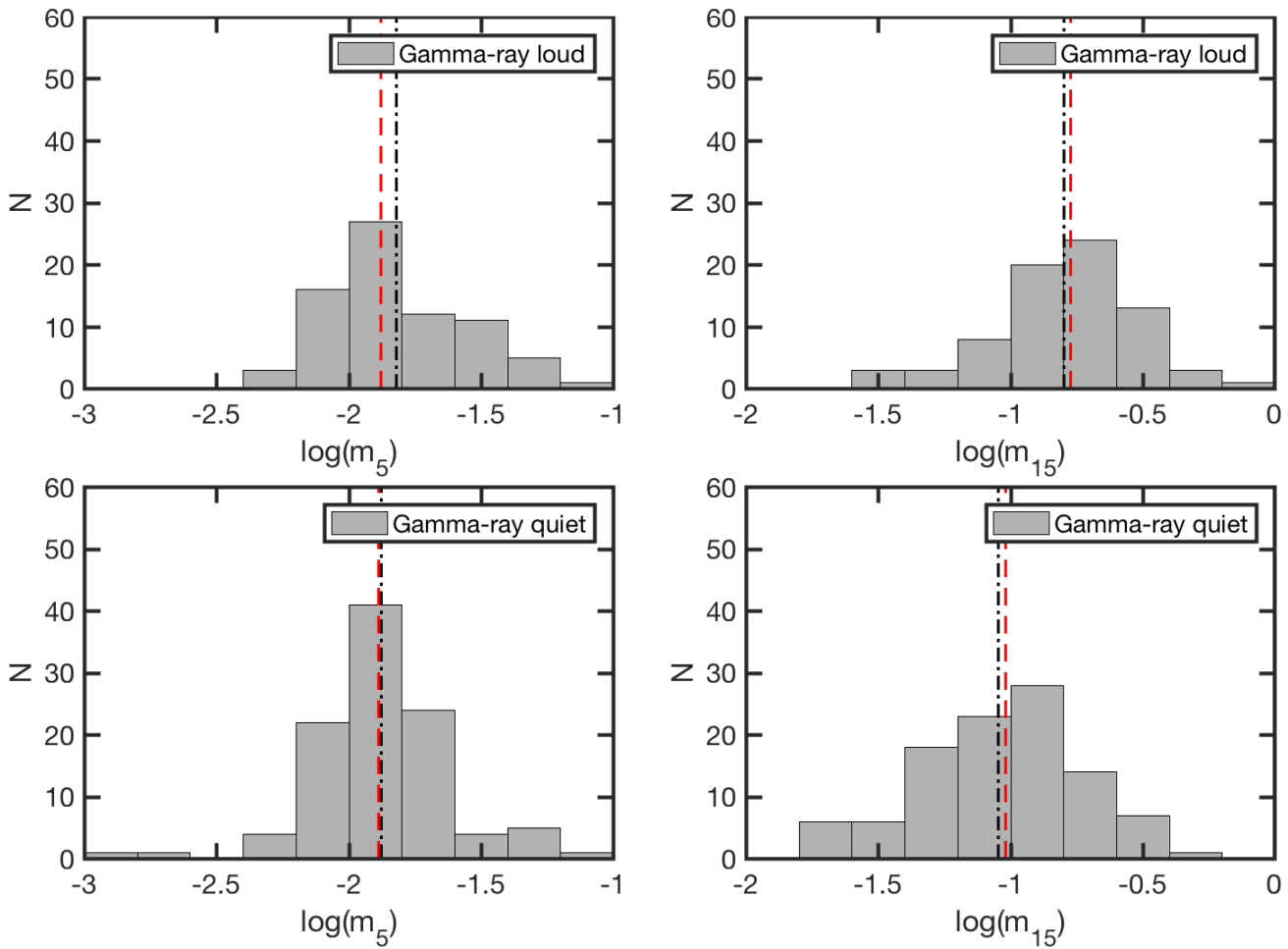}
	\end{center}
	\caption{{Distributions of the 5\,GHz modulation indices (left) and 15\,GHz modulation indices (right) for both the gamma-ray loud (\textit{Fermi}-detected, 2LAC, top) and gamma-ray quiet (no Fermi 2LAC detections, bottom) sources from the M $\cap$ R sample of 178 AGNs. The dashed (red) and dash-dotted (black) vertical lines in each panel show the median and mean values of the modulation indices, respectively. \label{vargammaray}}}
\end{figure*}

For our smaller M $\cap$ R sample of 178 sources, this strong connection between gamma-ray loudness (using 2LAC) and $m_{15}$ persists (Figure~\ref{vargammaray}, top and bottom right). The K-S test shows that the distribution of $m_{15}$ in the gamma-ray loud and gamma-ray quiet sub-samples are not drawn from the same parent population, with a $p$-value of $8.94 \times 10^{-7}$. 

However, we do not find a statistically significant connection between gamma-ray loudness and ISS amplitudes for this M $\cap$ R sample of 178 sources. The K-S test finds no significant difference ($p$-value of 0.160) between the distributions of $m_{5}$ in the gamma-ray loud and gamma-ray quiet sub-samples drawn from the M $\cap$ R sample (see also Figure~\ref{vargammaray}, top and bottom left). We use the 2LAC for identifying the gamma-ray loud sources, to compare our results directly with that published by \citet{richardsetal14}. 

\begin{figure*}
	\begin{center}
		\includegraphics[width=\textwidth]{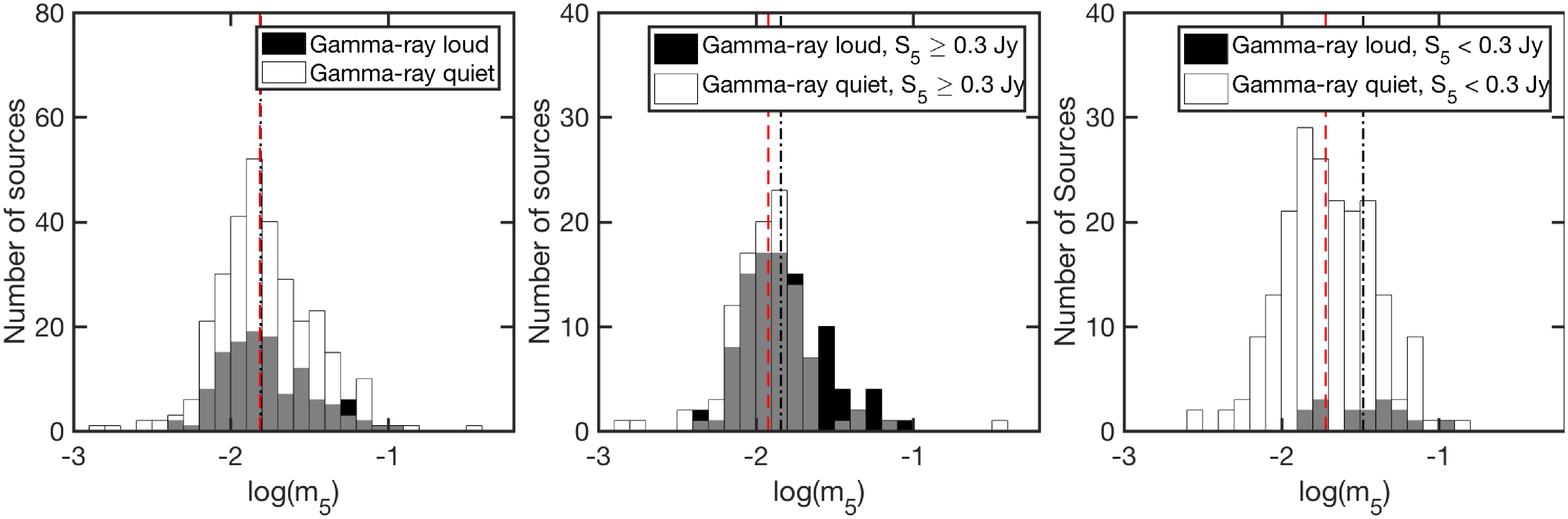}
	\end{center}
	\caption{{(Left) Distribution of the 5\,GHz modulation indices ($m_5$) for both the gamma-ray loud (with a \textit{Fermi} 3LAC association, black histogram) and gamma-ray quiet (no 3LAC association, white histogram) AGNs from the full MASIV sample of 424 sources. The middle and right panels show the same histograms, but for the high flux density ($S_5 \geq 0.3\,$Jy) and low flux density ($S_5 < 0.3\,$Jy) sources, respectively. The black dash-dotted and red dashed lines show the mean values of $m_5$ for the gamma-ray loud and gamma-ray quiet sources respectively. \label{gm3lac}}}
\end{figure*}

For the full MASIV sample of 424 sources, we also find no significant difference in the distributions of $m_5$ for the gamma-ray loud and gamma-ray quiet sources. This is true when we use both the 2LAC and also the Third LAT AGN Catalog (3LAC) to identify the gamma-ray loud sources; the K-S test gives a $p$-value of 0.976 when comparing the distributions of $m_5$ of the 3LAC-detected and non-detected samples. The distributions of $m_5$ for both 3LAC-detected and non-detected AGNs from the full MASIV sample are shown in the left panel of Figure~\ref{gm3lac}. 

However, when the 424 MASIV sources are separated into two groups based on their flux densities ($S_5 \geq 0.3\,$Jy and $S_5 < 0.3\,$Jy, based on the original MASIV classification of radio-strong and radio-weak sources), we find that the $m_5$ values of gamma-ray loud sources become significantly higher than that of the gamma-ray quiet sources! This is clearly seen in the middle and right panels of Figure~\ref{gm3lac}. The K-S test rejects the null hypothesis that the distributions of $m_5$ are the same for the gamma-ray loud and gamma-ray quiet samples. For the high flux density sources, the $p$-value is $7.2 \times 10^{-3}$, while for the low flux density sources, the $p$-value is $5.1 \times 10^{-3}$.

Why is there no significant difference in the $m_5$ distributions of gamma-ray loud and gamma-ray quiet sources from the full MASIV sample containing both the strong and weak flux density sources? This arises from the low fraction (7\%) of gamma-ray loud sources in the radio-weak sample (Figure~\ref{gm3lac}, right), as compared to the higher 50\% fraction of gamma-ray loud sources in the radio-strong sample (Figure~\ref{gm3lac}, middle), as expected due to the known radio-gamma-ray flux correlation in blazars \citep[e.g.][]{ackermannetal11a,max-moerbecketal14}. This selection bias, combined with the significantly higher $m_5$ of radio-weak sources relative to the radio-strong sources (Section~\ref{mutualflux}), dilutes the relationship between $m_5$ and gamma-ray loudness clearly seen when they are separated by flux density.

For the smaller M $\cap$ R sample of 178 sources, only 5 sources have 5\,GHz flux densities below 0.3\,Jy. Any dilution of the $m_5$ and gamma-ray loudness relationship would be negligible. Therefore, we can conclude that there is a weaker correspondence between the 5\,GHz variability amplitudes and gamma-ray loudness, compared to the much stronger relationship found between gamma-ray loudness and the 15\,GHz intrinsic variations. We attribute this weaker relationship to opacity effects. Relative to the 15\,GHz observations, the 5\,GHz observations probe regions downstream of the jet, and thus further from the gamma-ray emitting regions. 

\begin{figure}
	\begin{center}
		\includegraphics[width=\columnwidth]{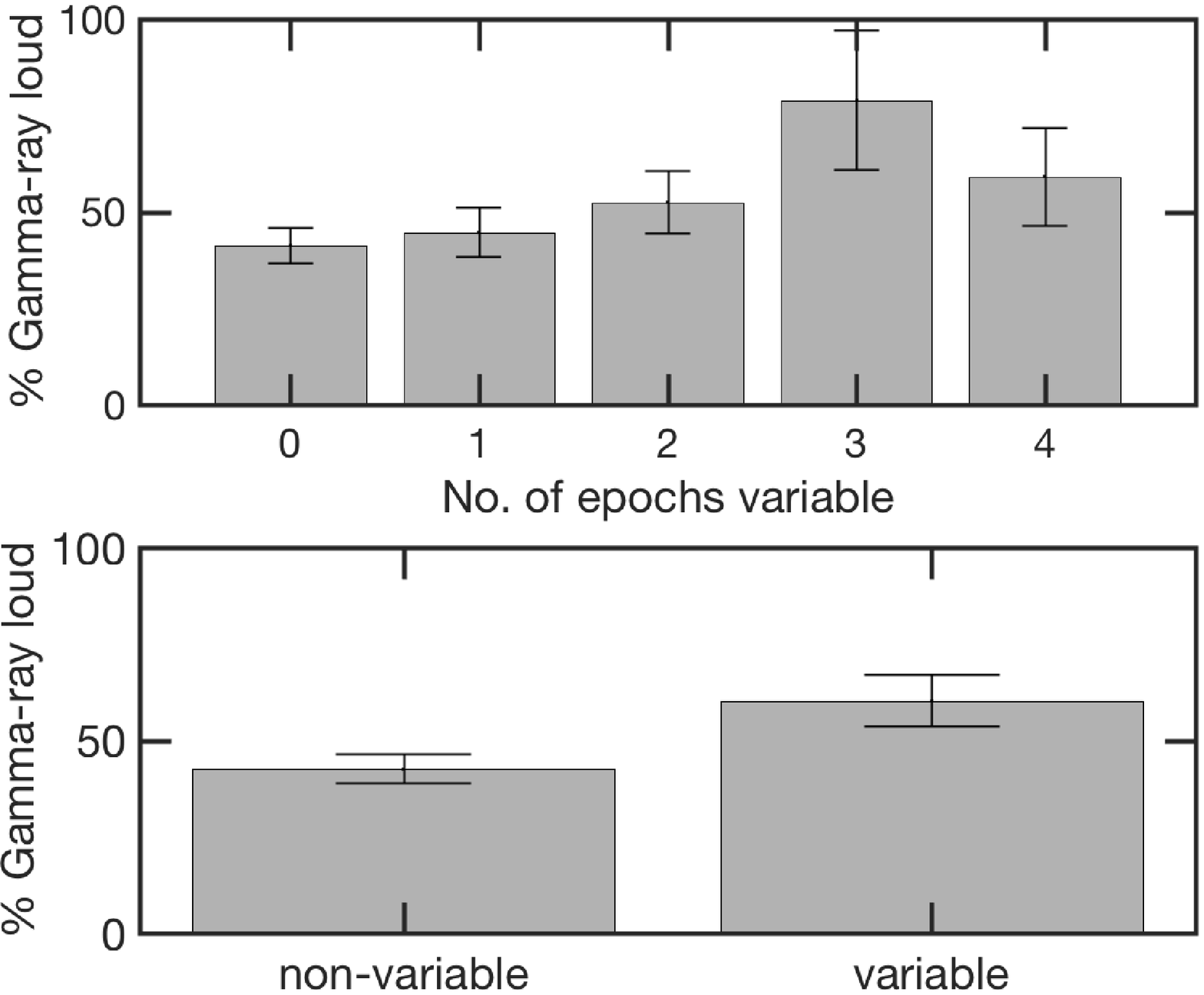}
	\end{center}
	\caption{{(Top) Fraction of sources in the MASIV Survey that are gamma-ray loud (associated with a \textit{Fermi}-detected 3LAC source), binned based on the number of MASIV Survey epochs in which the source was found to be scintillating. Only the high flux density ($S_5 \geq 0.3\,$Jy) sources are considered here. (Bottom) Same as the top panel, with the sources separated into two bins, one comprising 'non-variable' sources (scintillating in 0 or only 1 MASIV epoch) and another comprising 'variable' sources (scintillating in at least 2 MASIV epochs. The difference in fraction of gamma-ray loud sources in both bins are $< 3 \sigma$). The error bars are standards errors. \label{gm3lachifrac}}}
\end{figure}

For the AGNs found to be scintillating in 2 to 4 MASIV epochs, we find a slightly higher fraction of gamma-ray loud (\textit{Fermi} 3LAC detected) sources compared to AGNs scintillating in 1 or 0 epochs (Figure~\ref{gm3lachifrac}). This relationship between gamma-ray loudness and persistence of ISS is not significant ($< 2\sigma$), even if we separate the sources into only 2 bins, one for `variable' sources (those variable in at least 2 epochs) and another for `non-variable' sources (those variable in 0 or 1 epochs). If real, such behavior may be related to the observed connection between intrinsic variability amplitudes and the intermittency of ISS discussed in Section~\ref{intextepoch}. 

\section{Summary and Conclusions}\label{summary}

In this study, we examined the relationship between 5\,GHz intraday variability and the 15\,GHz flux variations of a sample of radio-selected compact AGNs, plus their dependences on the AGN intrinsic properties. Our main results are summarized as follows:

\begin{enumerate}
	
\item For the sample of 178 overlapping sources between the MASIV Survey and the OVRO blazar monitoring program, we confirm that the 5\,GHz variability amplitudes are dominated by ISS, while the 15\,GHz variations are dominated by intrinsic variability. 
	
\item We find a weak but statistically significant correspondence between the 5\,GHz ISS amplitudes and that of the 15\,GHz intrinsic variations, despite the observations being conducted more than $6$ years apart. 

\item All sources exhibiting the largest amplitude ISS at 5\,GHz ($m_5 \geq 0.02$) also show strong 15\,GHz intrinsic variability ($m_{15} \geq 0.1$). We argue that this relationship arises mainly due to the mutual dependence of $m_5$ and $m_{15}$ on the $\sim 100\, \mu$as scale core sizes of the sources; we observe a statistically significant dependence of both $m_5$ and $m_{15}$ on published 8 and 15\,GHz core sizes derived from Gaussian fits to VLBI visibility data.

\item The mutual dependences of $m_5$ and $m_{15}$ on source flux density, spectral index (as a measure of arcsec-scale core dominance), and redshift may be contributing factors, but are not as important as their $\sim 100\, \mu$as core size dependences, in establishing the observed relationship between intrinsic variability and ISS amplitudes. 

\item We found a population of sources that show strong intrinsic variations ($m_{15} \geq 0.1$), yet display only low level ISS ($m_5 \leq 0.02$). We attribute this to the nature of ISS that is also highly dependent on the properties of the line-of-sight scattering material in the ISM. Selecting only the $m_{15} \geq 0.1$ sources, we find that sources with high amplitude ISS ($m_5 \geq 0.02$) have significantly higher line-of-sight Galactic H$\alpha$ intensities than those with lower amplitude ISS. 

\item For the M $\cap$ R sample of 178 sources, the $m_5$ distributions of the gamma-ray loud and gamma-ray quiet AGNs are comparable, even though the gamma-ray loud AGNs have significantly higher $m_{15}$ than gamma-ray quiet AGNs. However, we discover that gamma-ray loud sources have typically higher $m_5$ than gamma-ray quiet sources when examining the full MASIV sample of sources with high and low flux density sources separated. The weaker relationship between $m_5$ and gamma-ray loudness (compared to the strong relationship found between $m_{15}$ and gamma-ray loudness) is likely due to opacity effects.

\item Persistent scintillators (i.e., those observed to be scintillating in 3 or 4 MASIV Survey epochs) are more likely to be intrinsically variable. This suggests that sources exhibiting strong intrinsic variability either contain a long-lived compact, scintillating component, or eject such compact components (e.g., via a flare) more regularly. 

\end{enumerate}

Our results showing the connection between AGN ISS and intrinsic variability demonstrate that both phenomena can be observed in the same population of sources in a radio variability survey; whether ISS or intrinsic variability dominates greatly depends on the observing frequency, cadence, and monitoring timespan of such a survey. The method of variability amplitude characterization is also important - by using the modulation index, for example, $m_{15}$ values derived by \citet{richardsetal11} are dominated by the largest inflections in the 4 year lightcurves, mostly intrinsic variations. 

Examining the Galactic and line-of-sight H$\alpha$ intensity dependence of the variability amplitudes provides the best way to distinguish between intrinsic and extrinsic variations for a large sample of sources. There is therefore a good case to be made for conducting all-sky or very wide-field variability surveys. On the other hand, single-field monitoring for variable sources will need to depend on other methods, such as selecting the appropriate observing frequency and cadence such that either ISS or intrinsic variability is suppressed. Alternatively, multi-frequency surveys can help in distinguishing between intrinsic and extrinsic variations through comparisons of observed variability characeristics with ISS models and Monte-carlo simulations, as demonstrated in Appendix~\ref{altexp}. Since both ISS and intrinsic variability depend on source flux density, spectral index, core dominance, and redshift, these properties cannot be used to distinguish between ISS and intrinsic variability in a large population of variable sources. 

Future variability surveys on present and upcoming radio telescopes will enable us to study and better determine the dominant mode of variability over a wider range of timescales and observing frequencies. For example, the new VLBI Global Observing System (VGOS) network will monitor International Celestial Reference Frame quasars over the whole sky, mainly for astrometric and geodetic applicatons \citep{petrachenkoetal13}. Operating at frequencies spanning 3 to 14\,GHz, the observed multi-frequency lightcurves and mas source structures will need to be analysed regularly, since they introduce systematic errors into geodetic measurements at the 1\,mm precision level \citep{shabalaetal14}. These data, based on coeval observations, present a significant improvement over the data we have used for our present study of the inter-relationships between ISS, intrinsic variability, and mas source structures. Through such studies, the ability to better distinguish between intrinsic and extrinsic radio variability of AGNs at different frequencies and timescales will pave the way to better understanding the underlying physical processes responsible for these variations, thus providing a powerful probe of AGN accretion and jet physics.

\section*{Acknowledgements}

The authors would like to thank the anonymous reviewer for insightful comments and suggestions to improve the manuscript. The authors are also grateful to Masanori Nakamura for helpful discussions. A part of this work was conducted while JYK was supported by a research grant (VKR023371) from Villumfonden. JYK gratefully acknowledges support from the Danish Council for Independent Research via grant no. DFF 4002-00275. The Dark Cosmology Centre is funded by the Danish National Research Foundation. MG thanks CIRA for the financial support and kind hospitality during his visit. This research has made use of published data from the OVRO 40-m monitoring program (Richards, J. L. et al. 2011, ApJS, 194, 29) which is supported in part by NASA grants NNX08AW31G, NNX11A043G, and NNX14AQ89G and NSF grants AST-0808050 and AST-1109911. The Wisconsin H$\alpha$ Mapper and its H$\alpha$ Sky Survey have been funded primarily by the National Science Foundation. The facility was designed and built with the help of the University of Wisconsin Graduate School, Physical Sciences Lab, and Space Astronomy Lab. NOAO staff at Kitt Peak and Cerro Tololo provided on-site support for its remote operation. The National Radio Astronomy Observatory is a facility of the National Science Foundation operated under cooperative agreement by Associated Universities, Inc.








\appendix

\section{Brief description of correlation tests used}\label{corrstat}

To determine the strength of correlation between 2 variables $X$ and $Y$, we used the Spearman rank correlation coefficient, $r_s$, defined as \citep[e.g.,][]{hollanderwolfe73}:
\begin{equation}
r_s = \dfrac{{\rm cov}(R_{X},R_{Y})}{\sigma_{R_{X}}\sigma_{R_{Y}}} \,\,\, , \label{spearman}
\end{equation}
where $R_{X}$ and $R_{Y}$ are rank variables of $X$ and $Y$, $\sigma_{R_{X}}$ and $\sigma_{R_{X}}$ the standard deviations of $R_{X}$ and $R_{Y}$, while $\rm cov$ is the covariance function. Tied ranks are adjusted for by using the averaged rank values. The Spearman correlation tests in our analyses were carried out using the \texttt{corr} function in the \textsc{MATLAB} software package. This function calculates $p$-values for the Spearman test using either the exact permutation distributions for small samples, or large-sample approximations where a normal distribution is assumed. For a two-tailed test, which we used for all our analyses, the $p$-value is estimated as twice that of the more significant of the two one-tailed $p$-values.

To account for confounding variables when testing for correlations, we used the \texttt{partialcorr} function in \textsc{MATLAB} to perform partial Spearman rank correlation tests. This function calculates the correlation coefficient between two variables $X$ and $Y$ (and corresponding $p$-value) while controlling for the effect of the variable $Z$ (or more variables). Linear least-square regressions between $X$ and $Z$, as well as between $Y$ and $Z$, are performed with $Z$ as the predictor; the residuals, $E_{XZ}$ and $E_{YZ}$ respectively, are calculated. The partial correlation coefficient between $X$ and $Y$ while controlling for $Z$ is then estimated as the Spearman's correlation coefficient between $E_{XZ}$ and $E_{YZ}$ following Equation~\ref{spearman}.

\section{Ruling out interpretation as correlated 5\,GHz and 15\,GHz ISS amplitudes}\label{altexp}

Here, we consider the possibility that the observed relationship between the 5\,GHz and 15\,GHz variability amplitudes arises from contamination of $m_{15}$ by ISS. 

Besides the non-dependence of $m_{15}$ on line-of-sight H$\alpha$ intensities (Section~\ref{intextvar}), another argument against ISS dominating the 15\,GHz modulation indices, is that the typical values of $m_{15}$ are about an order of magnitude larger than that of $m_{5}$. The mean and median values of $m_{15}$ are 0.146 and 0.122 respectively. On the other hand, the mean and median values of $m_{5}$ are lower, 0.017 and 0.013 respectively. ISS amplitudes at 15\,GHz are expected to be lower than that at 5\,GHz in the regime of weak ISS \citep[e.g.,][]{walker98}, which is the opposite of what we see.

\begin{figure*}
	\begin{center}
		\includegraphics[width=\textwidth]{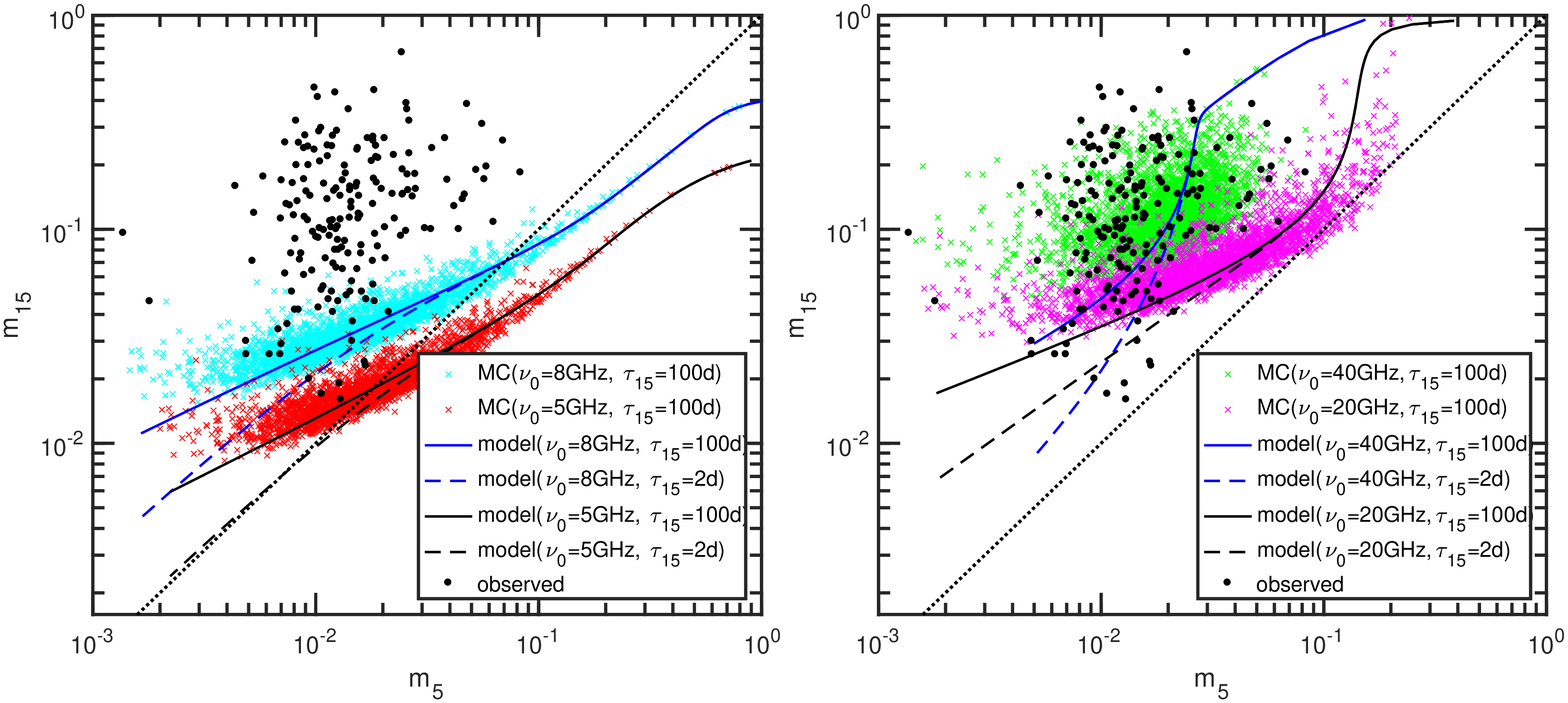}
	\end{center}
	\caption{{Model values of $m_{15}$ vs $m_{5}$ plotted for source angular sizes ranging from 1\,$\mu$as to 500\,$\mu$as at 5\,GHz (and assuming $\theta \propto \nu^{-1}$). The model values are estimated based on the \citet{goodmannarayan06} fitting formula for ISS, for cases where $\nu_0 = 5\,$GHz (black curves, left panel), $\nu_0 = 8\,$GHz (blue curves, left panel), $\nu_0 = 20\,$GHz (black curves, right panel) and $\nu_0 = 40\,$GHz (blue curves, right panel); $\nu_0$ is the transition frequency between weak and strong ISS. For all model curves, $m_5$ is derived from the model values of $D_{5}(\,\tau_{5} = 2d)$, while we show model $m_{15}$ values derived from both $D_{15}(\rm \tau_{15} = 2d)$ (dashed lines) and $D_{15}(\rm \tau_{15} = 100d)$ (solid lines). The diagonal black dotted lines in both panels represent the $x = y$ line. The models assume Kolmogorov turbulence, and that the scattering screen is located at a distance 500\,pc away, having a relative velocity of 30\,km\,s$^{-1}$. The colored scatter plots show simulated values of $m_{15}$ and $m_{5}$ for 3000 sources (in each instance of $\nu_0$), with Gaussian distributions of source angular sizes and scattering screen properties (Table~\ref{gaussian}). These simulated $m_{15}$ values are derived only using $D_{15}(\rm \tau = 100d)$. Observed values of $m_{15}$ and $m_{5}$ are also overlayed (black circles), with error bars removed for clarity. \label{varmcmodel}}}
\end{figure*}

Since our MASIV Survey sources are located typically at mid-Galactic latitudes where the observing frequencies of 5\,GHz and 15\,GHz are close to the transition frequency ($\nu_{0}$) between weak and strong scintillation, $m_{15}$ can in certain situations be larger than $m_{5}$ when $ 5\,{\rm GHz} \lesssim \nu_{0} \lesssim 15\,$GHz. This is demonstrated in the left panel of Figure~\ref{varmcmodel}, where we plot the model values of $m_{15}$ vs $m_{5}$ for source angular sizes of $\rm 1\,\mu as \leq \theta \leq 500\,\mu as$ at 5\,GHz for cases where $\nu_{0} = 8\,$GHz (blue lines) and $\nu_{0} = 5\,$GHz (black solid and dashed lines). In this model, we assume that the intrinsic source sizes have a frequency dependence of $\theta \propto \nu^{-1}$. We also assume a scattering screen with Kolmogorov turbulence located at a distance of 500\,pc with a relative velocity of 30\,km\,s$^{-1}$. The model values of $m_5$ are estimated from $D(\tau = 2\rm d)$,  following Equation~\ref{mod}. We show curves for both cases in which $m_{15}$ is derived from $D(\tau = 2\rm d)$ and $D(\tau = \rm 100d)$, the latter being more representative of the longer term 15\,GHz OVRO measurements where $D(\tau)$ has saturated. We calculated the structure functions using the \citet{goodmannarayan06} fitting function for ISS, which has the advantage of being also applicable to cases where the observing frequencies are comparable to $\nu_{0}$. In the left panel of Figure~\ref{varmcmodel}, $m_{15}$ is larger than $m_5$ in the lower left region of the model curves (corresponding to larger source sizes). This occurs when the source angular sizes are sufficiently large such that the ISS timescales are longer and the 5\,GHz $D(\tau)$ no longer saturates within a 2-day timescale, while the 15\,GHz $D(\tau)$ rises in amplitude (and saturates) more rapidly with $\tau$ than at 5\,GHz, due to the smaller source sizes at higher frequencies implemented in the model. However, in this scenario the ISS models still significantly underpredict $m_{15}$, for the range of $m_5$ values that are consistent with that observed. Following the Monte-Carlo method, we obtain simulated values of $m_{15}$ and $m_{5}$ for 3000 sources, calculated from the fitting fuction using randomly generated values of source angular sizes, scattering screen distances and scattering screen velocities assuming they have Gaussian distributions (their mean values and standard deviations are given in Table~\ref{gaussian}). The simulated values are shown as coloured scatter plots in the left panel of Figure~\ref{varmcmodel} for $\nu_0 = 8\,$GHz and $\nu_0 = 5\,$GHz. Clearly, the simulated values of $m_{15}$ are lower than that of the observed values. ISS at frequencies close to the boundary between weak and strong scattering cannot sufficiently account for the large amplitudes of $m_{15}$. 

\begin{table}
	\centering
	\caption{Mean and standard deviation ($\sigma$) of the Gaussian distribution for parameters used to generate simulated modulation indices of scintillating sources.
		\label{gaussian}}
	\begin{tabular}{lcc} 
		\hline 
		\hline
		Parameter	& Mean &  $\sigma$\\ 
		\hline
		Source FWHM angular size ($\mu$as)	& 150  &  50 \\ 
		Scattering screen distance (pc)	& 500 & 100 \\ 
		Scattering screen velocity (km\,s$^{-1}$) & 30 & 10 \\ 
		\hline 
	\end{tabular} 
\end{table}

In the regime of strong refractive ISS, we expect to observe even larger amplitude variations at 15\,GHz relative to that at 5\,GHz. Model values of $m_{15}$ vs $m_5$ for $\nu_{0} = 40\,$GHz (blue lines) and $\nu_{0} = 20\,$GHz (black solid and dashed lines) are shown in the right panel of Figure~\ref{varmcmodel}, derived using the same method and assumptions as that for the model estimates in the left panel. Colored scatter plots show 3000 simulated values of $m_{15}$ and $m_5$ derived using the same distributions of source sizes and scattering screen properties. Although the observed values of $m_{15}$ vs $m_5$ are comparable to that of the simulated values for the case where $\nu_{0} = 40\,$GHz, it is unlikely that $\nu_0$ has such high values for the majority of the sources in our sample. Transition frequencies of $\nu_0 \gtrsim 20$\,GHz are expected at Galactic latitudes of $\lvert b \rvert \lesssim 15^{\circ}$ \citep{walker98,walker01}, while the MASIV sources have typical Galactic latitudes of $\lvert b \rvert \gtrsim 20^{\circ}$. 

Furthermore, to scintillate at such high amplitudes at 15\,GHz, even in the strong refractive ISS regime, the sources would need to be sufficiently compact such that the scintillation timescales would only be of order a few hours. As an example, we consider the limiting case where a large fraction of our sources are observed through sufficiently low Galactic latitudes (or high line-of-sight H$\alpha$ intensities) such that $\nu_0 \sim 20$\,GHz. In this scenario, the sources would exhibit strong refractive ISS at both 5\,GHz and 15\,GHz. To exhibit the 10\% to 40\% flux variations at 15\,GHz as observed in 60\% of our M $\cap$ R sample of AGNs, these sources would need to be very compact ($\theta \lesssim 50\,\mu$as). However, such compact sources would be scintillating at characteristic timescales of $\lesssim 2$\,hrs, which would place them in the category of `extreme scintillators'. Such a scenario is highly unlikely since extreme scintillators are very rare, with only a handful discovered so far \citep{kedziora-chudczeretal97, dennett-thorpedebruyn00, bignalletal03}. Only 1 of the MASIV Survey sources (J1819+345) displayed such extreme scintillation at 5\,GHz. 

Another very obvious discrepancy between the simulated ISS variability amplitudes and the observed values in Figure~\ref{varmcmodel} is this: ISS amplitudes of a sample of sources with sight-lines covering a wide range of Galactic latitudes (i.e., with $\nu_0$ ranging from 5 to 40\,GHz), are expected to also populate the lower right quadrant of Figure~\ref{varmcmodel}. In other words, there should be a population of sources that scintillate at amplitudes of $m_5 \geq 0.02$ \textit{and} that exhibit 15\,GHz ISS amplitudes of $m_{15} \leq 0.1$. However, this is not what we see in the observational data. This strongly suggests that, for these long, 4-year timespan 15\,GHz OVRO observations, the low-level ISS-induced variations on shorter timescales are neligible when compared to the higher amplitude intrinsic variations characterized by $m_{15}$. 

Although $m_{15}$ is monitored over a much longer timespan relative to $m_{5}$, potentially allowing larger amplitude scintillations over longer timescales to be observed in the 15\,GHz OVRO lightcurves, we argue that this scenario is unlikely. From Figure~\ref{varmcmodel}, we see that the model curves for cases where $m_{15}$ is derived from $D(\tau = 2\rm d)$ is comparable to that where $m_{15}$ is derived from $D(\tau = 100\rm d)$, for the most variable $m_{15} \gtrsim 0.1$ sources. This implies that the ISS timescales are less than two days, such that the structure function amplitudes saturate by $\tau \lesssim 2\,\rm d$ for the most compact $\lesssim 50\,\mu$as sources. For more extended source components, the scintillation timescales can increase such that $m_{15}$  appears to be much larger than $m_5$ due to the longer observing span at 15\,GHz relative to that at 5\,GHz. The 5\,GHz structure function amplitudes in this case have yet to saturate since the observing period is less than the characteristic scintillation timescale. However, the scintillation amplitudes of such extended source components would be small ($m_{15} \lesssim 0.1$), and cannot explain the $m_{15} >0.1$ observed in 60\% of the M $\cap$ R sources. 

The models and simulations also demonstrate that the modulation index ratios, $R_{15}^5 = m_5/m_{15}$, should decrease with increasing values of $\nu_0$. Since the line-of-sight H$\alpha$ intensities function as a reasonable proxy of $\nu_0$, this predicts that we should observe an anti-correlation between $R_{15}^5$ and H$\alpha$ intensities, if ISS indeed dominates the variability at both frequencies. In fact, such a dependence of the ratio of ISS modulation indices on line-of-sight H$\alpha$ intensities has been observed in simultaneous 5\,GHz and 8\,GHz intraday variability surveys \citep{koayetal12}. For sight-lines with high H$\alpha$ intensities, sources are more likely to be scintillating in the strong scattering regime, leading to typically lower values of $R_{15}^5$. For our current dataset, instead of finding a negative correlation between $R_{15}^5$ and H$\alpha$ intensities as expected for ISS, we find a positive correlation ($\tau = 0.123$, $p = 0.017$); $R_{15}^5$ increases with increasing H$\alpha$ intensity. This is likely due to the fact that only $m_{5}$ increases while $m_{15}$ remains relatively constant with increasing H$\alpha$ intensities.

Ultimately, the 15\,GHz lightcurves need to be reexamined to determine the extent to which $m_{15}$ may be contaminated by ISS. By using structure function analyses, we can determine if the dependence of $m_{15}$ on H$\alpha$ intensity increases with decreasing timescales. The OVRO 15\,GHz lightcurves can also be examined for signatures of annual cycles in the variability timescales, a strong evidence for ISS. These are beyond the scope of our present work. Regardless, the arguments we have laid out in this Appendix strongly suggest that the observed correlation between $m_5$ and $m_{15}$ demonstrates a relationship between intra-day ISS at 5\,GHz and longer timescale intrinsic variations at 15\,GHz, as opposed to correlated ISS amplitudes at both frequencies. 

\bsp	
\label{lastpage}
\end{document}